\documentclass{article}

\usepackage{arxiv}

\usepackage[utf8]{inputenc} 
\usepackage[T1]{fontenc}    
\usepackage[hidelinks]{hyperref}       
\usepackage{url}            
\usepackage{booktabs}       
\usepackage{amsfonts}       
\usepackage{nicefrac}       
\usepackage{microtype}      
\usepackage{lipsum}		
\usepackage[dvipdfmx]{graphicx}
\usepackage{natbib}
\usepackage{doi}

\usepackage{bm}
\usepackage{amsmath}
\usepackage{mathtools}
\usepackage{enumerate}
\usepackage{color}

\newtheorem{Lem}{Lemma}

\newtheorem{Thm}{Theorem}

\newcommand{\df}{\mathop{\rm df}}
\newcommand{\diag}{\mathop{\rm diag}}
\newcommand{\rank}{\mathop{\rm rank}}

\newcommand{\tr}{\mathop{\rm tr}}

\newcommand{\E}{\mathop{\rm E}\nolimits}

\newcommand{\MSE}{\mathop{\rm MSE}\nolimits}

\newcommand{\RMSE}{\mathop{\rm RMSE}\nolimits}

\newcommand{\Var}{\mathop{\rm Var}\nolimits}

\renewcommand{\Pr}{\mathop{\rm P}}

\renewcommand{\vec}[1]{{\bm{#1}}}
\def\0{\vec{0}}
\def\1{\vec{1}}

\def\vb{\vec{b}}

\def\vx{\vec{x}}
\def\vy{\vec{y}}
\def\vz{\vec{z}}

\def\vA{\vec{A}}
\def\vB{\vec{B}}

\def\vD{\vec{D}}

\def\vH{\vec{H}}
\def\vI{\vec{I}}

\def\vM{\vec{M}}

\def\vO{\vec{O}}
\def\vP{\vec{P}}

\def\vU{\vec{U}}
\def\vV{\vec{V}}

\def\vX{\vec{X}}

\def\vZ{\vec{Z}}

\def\cA{\mathcal{A}}

\def\bN{\mathbb{N}}

\def\vbeta{\vec{\beta}}
\def\vgamma{\vec{\gamma}}

\def\veta{\vec{\eta}}
\def\vtheta{\vec{\theta}}

\def\vxi{\vec{\xi}}

\def\vPsi{\vec{\Psi}}

\title{KOO Method--based Consistent Clustering \\ for Group-wise Linear Regression \\ with Graph Structure}

\author{ 
  \href{https://orcid.org/0000-0001-8727-3631}{\includegraphics[scale=0.06]{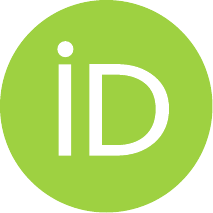}\hspace{1mm}Mineaki Ohishi}\thanks{
    Corresponding author 
  } \\
  Center for Data-driven Science and Artificial Intelligence \\
  Tohoku University\\
  Sendai, Japan \\
  \texttt{mineaki.ohishi.a4@tohoku.ac.jp} \\
  \And
  Ryoya Oda \\
  Graduate School of Advanced Science and Engineering\\
  Hiroshima University\\
  Higashi-Hiroshima, Japan 
}

\renewcommand{\headeright}{}
\renewcommand{\undertitle}{}
\renewcommand{\shorttitle}{Consistent Clustering for Group-wise Linear Regression}

\begin{document}
\maketitle

\begin{abstract}
  The kick-one-out (KOO) method is a variable selection method based on a model selection criterion.
  The method is very simple, and yet it has consistency in variable selection under a high-dimensional asymptotic framework with a specific model selection criterion.
  This paper proposes the join-two-together (JTT) method, which is a clustering method based on the KOO method for group-wise linear regression with graph structure.
 The JTT method formulates the clustering problem as an edge selection problem for a graph and determines whether to select each edge based on the KOO method.
  We can employ network Lasso to perform such a clustering.
  However, network Lasso is somewhat cumbersome because there is no good algorithm for solving the associated optimization problem and the tuning is complicated.
  Therefore, by deriving a model selection criterion such that the JTT method has consistency in clustering under a high-dimensional asymptotic framework, we propose a simple yet powerful method that outperforms network Lasso.
\end{abstract}


\section{Introduction}

For $m$ groups, suppose we have a dataset $(\vy_j, \vX_j)\ (j \in \{ 1, \ldots, m \})$, where $\vy_j$ is an $n_j$-dimensional vector of a response variable, $\vX_j$ is an $n_j \times p$ matrix of explanatory variables satisfying $\rank (\vX_j) = p \le n_j$, and $n_j$ is the sample size of the $j$th group.
For such a dataset, we then assume the following group-wise linear regression model: 
\begin{align}
\label{model}
  \vy_j \sim N_{n_j} (\vX_j \vbeta_j, \sigma^2 \vI_{n_j}),
\end{align}
where $\vbeta_j$ is a $p$-dimensional vector of regression coefficients, $\sigma^2$ is an error variance satisfying $\sigma > 0$, and $\vy_1, \ldots, \vy_m$ are mutually independent.
Furthermore, we assume $N - 4 > 0$, where $N = n - mp$ and $n = \sum_{j=1}^m n_j$.
A simple method for estimating $\vbeta_1, \ldots, \vbeta_m$ is to apply the ordinary least squares (OLS) or maximum likelihood estimation method group-wise.
The group-wise estimator of $\vbeta_j\ (j \in \{ 1, \ldots, m \})$ obtained from either method is as follows:
\begin{align}
\label{gwLSE}
  \hat{\vbeta}_j
    = \vM_j^{-1} \vX_j' \vy_j, \quad
  \vM_j
    = \vX_j' \vX_j.
\end{align}
If $m$ groups have relationships with each other, it would be desirable to utilize an estimation method that considers those relationships rather than the simple group-wise estimation method given above.
In the present paper, we assume that the relationships among the groups can be formulated as a graph.
That is, we assume that the relationships are given by a graph $(V, E)$ with a vertex set $V = \{ 1, \ldots, m \}$ and an edge set $E\ (\subset V \times V)$ satisfying $(k, \ell) \in E \Rightarrow k < \ell$.
For example, by regarding houses or areas as groups, we can define $E$ in terms of regional adjacency (e.g., \citealp{Hallac-2015, Ohishi-2025}).
If $(k, \ell) \in E$, the $k$th and $\ell$th groups are related, and if $(k, \ell) \notin E$, they are not related.
As a method for estimating $\vbeta_1, \ldots, \vbeta_m$ under such a graph structure, we can consider a penalized estimation method with the penalty based on differences of regression coefficient vectors for related groups, i.e., $\vbeta_k - \vbeta_\ell\ ((k, \ell) \in E)$.
Specifically, we can use network Lasso (\citealp{Hallac-2015}) to obtain estimators satisfying $\vbeta_k = \vbeta_\ell$ exactly for some $(k, \ell) \in E$.
The network Lasso estimator is obtained by minimizing the following penalized residual sum of squares:
\begin{align}
\label{nwlasso}
  \sum_{j=1}^m \| \vy_j - \vX_j \vbeta_j \|^2
    + \lambda \sum_{(k, \ell) \in E} w_{k \ell} \| \vbeta_k - \vbeta_\ell \|,
\end{align}
where $\lambda\ (\ge 0)$ is a tuning parameter adjusting the strength of the penalty against the model fitting and $w_{k \ell}\ (> 0)$ is a penalty weight expressing the strength of the relationship between the $k$th and $\ell$th groups.
Network Lasso can perform an estimation that considers the relationships among the groups by shrinking $\vbeta_k - \vbeta_\ell$ based on a network structure. 
Notably, it allows $\| \vbeta_k - \vbeta_\ell \|$ to be exactly zero.
This implies that $\vbeta_k$ and $\vbeta_\ell$ can be estimated as exactly equal. 
In this case, we can perform clustering by interpreting that the $k$th and $\ell$th groups belong to the same cluster.
However, network Lasso is somewhat cumbersome in practice.
One of its problems is the optimization method.
If clustering is not required, then the optimization method is not a problem.
For example, we can adopt $\| \vbeta_k - \vbeta_\ell \|^2$ instead of $\| \vbeta_k - \vbeta_\ell \|$ in \eqref{nwlasso}.
In this case, we can perform an estimation based on a network structure and the estimator can be obtained in closed form.
However, since the network Lasso estimator cannot be obtained in closed form, its optimization method is important.
\cite{Hallac-2015} proposed an algorithm based on the alternating direction method of multipliers (ADMM; \citealp{Boyd-2011}) to minimize \eqref{nwlasso}.
ADMM is a very popular method because it is highly versatile and has good theoretical properties, e.g., optimality and convergence.
On the other hand, it too has some problems in practice.
We focus here on two problems: the convergence of the algorithm is slow and $\vbeta_k$ and $\vbeta_\ell$ cannot be estimated as numerically and exactly equal for network Lasso.
The latter in particular is serious because clustering is one purpose of applying  network Lasso.
In order to overcome these problems in ADMM, \cite{Ohishi-2025} proposed an algorithm based on the block-wise coordinate descent method (BCDM).
BCDM can estimate $\vbeta_k$ and $\vbeta_\ell$ as numerically and exactly equal and \cite{Ohishi-2025} reported  that BCDM is superior to ADMM in terms of optimization speed and minimization accuracy.
However, since the penalty is not separable with respect to $\vbeta_j$, BCDM is not guaranteed to have the desired theoretical properties, e.g., optimality and convergence.
As mentioned above, for network Lasso, there is not an algorithm with both theoretical and practical goodness.
In addition, since network Lasso requires the selection of $\lambda$ and $w_{k \ell}$, network Lasso cannot be said to be easy to use.

The present paper focuses on clustering based on a graph and considers an approach that is completely different from continuous optimization methods such as network Lasso.
Specifically, we focus on the kick-one-out (KOO) method (\citealp{Zhao-1986, Nishii-1988}), which is a variable selection method (the name ``kick-one-out'' was given by \citealp{Bai-2018tr}).
The KOO method determines whether to select the $j$th variable by comparing a model which has all $p$ variables with a model excluding only the $j$th variable and the goodness for each model is evaluated by a model selection criterion, e.g., the Akaike information criterion (\citealp{Akaike1973}) or the $C_p$ criterion (\citealp{Mallows1973}).
Hence, the KOO method requires only $p$ calculations and is feasible in a large number of variables.
Despite the KOO method being very simple, as just described, it has consistency in variable selection.
For example, for the problem of selecting explanatory variables in multivariate linear regression, \cite{OdaYanagihara2020, OdaYanagihara2021} revealed classes of the generalized $C_p$ ($GC_p$) criterion (\citealp{Atkinson1980}) and generalized information criterion (\citealp{Nishii1984}), respectively, such that the KOO method has consistency in variable selection under a high-dimensional asymptotic framework, and proposed specific criteria.
Furthermore, \cite{OdaYanagihara2020} reported that the KOO method is superior to adaptive group Lasso (\citealp{WangLeng2008}), which is a continuous optimization method, in terms of performance of variable selection and calculation time.
Inspired by the KOO method, we propose the join-two-together (JTT) method, which performs clustering for group-wise linear regression.
The JTT method formulates the clustering problem as the problem of selecting  the pairs $(k, \ell)\ (\in E)$ that should belong to the same cluster.
In other words, the clustering problem is formulated as the problem of selecting edges $(k, \ell)$, where if $(k, \ell)$ is selected, the $k$th and $\ell$th groups are interpreted as belonging to the same cluster.
Specifically, the JTT method determines whether to select the edge $(k, \ell)$ (i.e., whether $\vbeta_k = \vbeta_\ell$) by comparing a model in which all regression coefficient vectors are different with a model in which $\vbeta_k = \vbeta_\ell$ only for $(k, \ell)$.
When $(V, E)$ is a complete graph, $\# (E)$ achieves the maximum and is ${}_m C_2 = m (m - 1) / 2$.
Hence, the JTT method requires a calculation of $O (m^2)$.
In the present paper, the $GC_p$ criterion is employed to evaluate the goodness of a model.
Based on \cite{OdaYanagihara2020}, we reveal a class of the $GC_p$ criterion such that the JTT method has consistency in the edge selection under a high-dimensional asymptotic framework and propose a specific criterion.
Moreover, we show that the JTT method is superior to network Lasso through numerical studies.

The remainder of the paper is organized as follows. 
In Section 2, we describe the JTT method and give the asymptotic framework and assumptions to discuss consistency.
In Section 3, we describe the main results: consistency of the JTT method and an estimation method after clustering.
In Sections 4 and 5, we numerically compare the JTT method with network Lasso, using simulation data and real data.
Section 6 concludes the paper.
Technical details are provided in the Appendices.

\section{Preliminaries}

We first describe the models and framework of the JTT method.
Let $E_\ast$ be a set of true edges, and define $m_\ast$ as the number of connected components in the graph $(V, E_\ast)$ and $V_i^\ast\ \ (i \in \{ 1, \ldots, m_\ast \})$ as the vertex set of the $i$th connected component.
That is, $m_\ast$ is the number of true clusters and $V_1^\ast, \ldots, V_{m_\ast}^\ast$ are the nonempty and mutually exclusive sets expressing true clusters, where $V = \cup_{i=1}^{m_\ast} V_i^\ast$.
The $E_\ast$ and $V_i^\ast$ have the following relationships:
\begin{align*}
  &\forall k, \ell \in V_i^\ast,\ (k, \ell) \in E \Longrightarrow (k, \ell) \in E_\ast, \quad
  \forall (k, \ell) \in E_\ast,\ \exists! i \in \{ 1, \ldots, m_\ast \}\ s.t.\ k, \ell \in V_i^\ast.
\end{align*}
Then, we define the true model as
\begin{align*}
  \vy_j \sim N_{n_j} (\vX_j \vbeta_j^\ast, \sigma_\ast^2 \vI_{n_j})\ (j \in V), \quad
  \vbeta_j^\ast = \vxi_i^\ast\ (j \in V_i^\ast;\ i \in \{ 1, \ldots, m_\ast \}),
\end{align*}
where $\vbeta_j^\ast$ is the $p$-dimensional vector of the true regression coefficients for the $j$th group, $\sigma_\ast^2$ is the true error variance satisfying $\sigma_\ast > 0$, and $\vxi_i^\ast$ is the $p$-dimensional vector of common regression coefficients for the groups in $V_i^\ast$.
Thus, $\vxi_i^\ast$ expresses the relationships among the groups, and the mean structure of the true model is equal within the same cluster.
We write $\vbeta_k^\ast = \vbeta_\ell^\ast = \vxi_{k \ell}^\ast$ when $(k, \ell) \in E_\ast$.
For example, when $m = 5$ and $E_\ast = \{ (1, 2), (1, 3), (4, 5) \}$, we have $m_\ast = 2$, $V_1^\ast = \{ 1, 2, 3 \}$, and $V_2^\ast = \{ 4, 5 \}$.
Furthermore, we have $\vbeta_1^\ast = \vbeta_2^\ast = \vxi_{12}^\ast = \vxi_1^\ast$.
For the true model, we define the base model as \eqref{model}, in which $\vbeta_1, \ldots, \vbeta_m$ are all different and a candidate model as a model with $\vbeta_k = \vbeta_\ell = \vxi_{k \ell}$ only for $(k, \ell) \in E$.
The candidate model for $(k, \ell) \in E$ is given by
\begin{align*}
  \vy_j \sim \begin{dcases}
      N_{n_j} (\vX_j \vxi_{k \ell}, \sigma_{k \ell}^2 \vI_{n_j}) 
        & (j \in \{ k, \ell \}) \\
      N_{n_j} (\vX_j \vbeta_j, \sigma_{k \ell}^2 \vI_{n_j}) 
        & (j \in V_{k \ell})
    \end{dcases},
\end{align*}
where $\sigma_{k \ell}^2$ is an error variance satisfying $\sigma_{k\ell} > 0$ and $V_{k \ell} = V \backslash \{ k, \ell \}$.
The group-wise OLS estimator of $\vbeta_j$ for the candidate model is given by
\begin{align*}
  \hat{\vbeta}_j^{(k \ell)} &= \begin{dcases}
      \vM_{k \ell}^{-1} \vX_{k \ell}' \vy_{k \ell}
        & (j \in \{ k, \ell \}) \\
      \hat{\vbeta}_j
        & (j \in V_{k \ell})
    \end{dcases}, \quad
  \vM_{k \ell}
    = \vX_{k \ell}' \vX_{k \ell}, \quad
  \vX_{k \ell} = \begin{pmatrix}
      \vX_k \\ \vX_\ell
    \end{pmatrix}, \quad
  \vy_{k \ell} = \begin{pmatrix} 
      \vy_k \\ \vy_\ell
    \end{pmatrix},
\end{align*}
where $\hat{\vbeta}_j$ is the group-wise OLS estimator of $\vbeta_j$ for the base model given in \eqref{gwLSE}.
Furthermore, we define an unbiased estimator $s^2$ of the error variance in the base model and projection matrices $\vP_j$ and $\vP_{k \ell}$ as
\begin{align*}
  s^2
    &= \dfrac{1}{N} \sum_{j=1}^m \vy_j' (\vI_{n_j} - \vP_j) \vy_j, \quad
  \vP_j
    = \vX_j \vM_j^{-1} \vX_j', \quad
  \vP_{k \ell}
    = \vX_{k \ell} \vM_{k \ell}^{-1} \vX_{k \ell}'.
\end{align*}
Then, the $GC_p$ criteria $GC_p^{(0)}$ and $GC_p^{(k \ell)}$ for the base model and the candidate model, respectively, are defined by
\begin{equation}
\label{GCp}
\begin{split}
  GC_p^{(0)} (\alpha)
    &= \dfrac{1}{s^2} \sum_{j=1}^m \vy_j' (\vI_{n_j} - \vP_j) \vy_j + \alpha m p, \\
  GC_p^{(k\ell)} (\alpha)
    &= \dfrac{1}{s^2} \left\{
           \vy_{k \ell}' (\vI_{n_{k \ell}} - \vP_{k \ell}) \vy_{k \ell}
             + \sum_{j \in V_{k \ell}} \vy_j' (\vI_{n_j} - \vP_j) \vy_j
         \right\} + \alpha (m-1) p,
\end{split}
\end{equation}
where $\alpha\ (> 0)$ is a parameter adjusting the strength of the penalty against the model fitting and $n_{k \ell} = n_k + n_\ell$.
The $GC_p$ criterion expresses a specific criterion by giving the $\alpha$ value, e.g., $\alpha=2$ expresses the $C_p$ criterion.
Under the models described above, we define the optimal edge set $\hat{E} (\alpha)$ selected by the JTT method based on the $GC_p$ criterion as
\begin{align}
\label{Ehat}
  \hat{E} (\alpha)
    = \left\{
         (k, \ell) \in E \mid GC_p^{(k \ell)} (\alpha) \le GC_p^{(0)} (\alpha)
       \right\}.
\end{align}
When the goodness of a model is evaluated by a model selection criterion, a model with a smaller value can be interpreted as being a better model.
In our case, we adopt $\vbeta_k = \vbeta_\ell$ satisfying $GC_p^{(k \ell)} (\alpha) \le GC_p^{(0)} (\alpha)$ for $(k, \ell) \in E$.
Note that the name ``kick-one-out'' is derived from considering a model from which only one variable is excluded as a candidate model.
Following this idea, we have named the proposed method the ``join-two-together'' method, since a model in which two groups are joined together is considered as a candidate model.
Let $q = \# (E)$. 
Then, $q$ achieves its maximum when $(V, E)$ is a complete graph.
Hence, the JTT method requires a calculation size of $q = O (m^2)$, since $q \le {}_m C_2 = m (m-1) / 2$.
The purpose of this paper is to present a condition on $\alpha$ such that the JTT method has consistency in the edge selection, i.e., the probability of $\hat{E} (\alpha) = E_\ast$ converges to 1.

Next, we provide an asymptotic framework and some assumptions for discussing the consistency.
Let $n_0 = \min_{j \in V} n_j$.
Herein, we consider consistency under the following high-dimensional asymptotic framework:
\begin{align}
\label{AFW}
  n_0 \rightarrow \infty, \quad
  \dfrac{p}{n_0} \rightarrow p_0 \in [0, 1), \quad
  \dfrac{m}{n_0} \rightarrow m_0 \in [0, \infty).
\end{align}
Under this asymptotic framework, $n_0$ always diverges to infinity.
In contrast, $p$, $m$, and $m_\ast$ can be fixed or diverge to infinity at a speed comparable to or slower than $n_0$.
We make the following three assumptions in preparation.
\begin{enumerate}[{Assumption A}1.]
\item
  $E_\ast \subseteq E$.
\item
  There exists $c_1 > 0$ such that
  \begin{align*}
    n_0^{-1} \min_{(k, \ell) \notin E_\ast} \lambda_{\min} (\vM_k \vM_{k \ell}^{-1} \vM_\ell) \ge c_1,
  \end{align*}
  where $\lambda_{\min} (\vA)$ is the minimum eigenvalue of a square matrix $\vA$.
\item
  There exist $c_2 > 0$ and $c_3 \ge 1/2$ such that
  \begin{align*}
    n_0^{1-c_3} \min_{(k, \ell) \notin E_\ast} \| \vbeta_k^\ast - \vbeta_\ell^\ast \|^2 / \sigma_\ast^2 \ge c_2.
  \end{align*}
\end{enumerate}
Assumption A1 is absolutely necessary for considering consistency.
Assumption A2 may seem unusual.
For example, the following might be considered a standard assumption.
\begin{enumerate}[{Assumption A2'}.]
\item
  There exists $d > 0$ such that
  \begin{align*}
    \min_{j \in V} n_j^{-1} \lambda_{\min} (\vM_j) \ge d.
  \end{align*}
\end{enumerate}
Actually, Assumption A2' is a sufficient condition for Assumption A2.
Hence, Assumption A2 is a relatively weak assumption.
Assumption A3 is a weak assumption for the true parameters.
Since it is necessary to determine $\vbeta_k \neq \vbeta_\ell$ for $(k, \ell) \notin E_\ast$, it would be desirable that $\| \vbeta_k^\ast - \vbeta_\ell^\ast \|^2$ be sufficiently large.
However, Assumption A3 allows $\| \vbeta_k^\ast - \vbeta_\ell^\ast \|^2 / \sigma_\ast^2$ to converge to 0.

Lastly, the asymptotic behavior of a non-centrality parameter is important in model selection.
The non-centrality parameter for $(k, \ell) \in E$ is given by
\begin{align*}
  \delta_{k \ell} 
    = {\veta_{k \ell}^\ast}' (\vI_{n_{k \ell}} - \vP_{k \ell}) \veta_{k \ell}^\ast / \sigma_\ast^2, \quad
  \veta_{k \ell}^\ast = \begin{dcases}
      \vX_{k \ell} \vxi_{k \ell}^\ast
        & ((k, \ell) \in E_\ast) \\
      \begin{pmatrix}
        \vX_k \vbeta_k^\ast \\ \vX_\ell \vbeta_\ell^\ast
      \end{pmatrix}
        & ((k, \ell) \notin E_\ast)
    \end{dcases}.
\end{align*}
Since $\vP_{k \ell} \vX_{k \ell} = \vX_{k \ell}$, we know that $\delta_{k \ell} = 0$ holds when $(k, \ell) \in E_\ast$ and $\delta_{k \ell} > 0$ holds when $(k, \ell) \notin E_\ast$.
Regarding $\delta_{k \ell}$, we have the following inequality under Assumptions A1--A3 (the proof is given in Appendix~\ref{ap delmin}):
\begin{align}
\label{delmin}
  n_0^{-c_3} \delta_{\min} \ge c_1 c_2, \quad
  \delta_{\min} = \min_{(k, \ell) \notin E_\ast} \delta_{k \ell}.
\end{align}

\section{Main results}

\subsection{Consistency}

We define a class of $\alpha$ of $GC_p$ criteria in \eqref{GCp} as 
\begin{align}
\label{class}
  \cA
    = \left\{
         \dfrac{N}{N - 2} + \beta\ \middle|\ 
           \beta > 0,\ 
           \dfrac{\beta p^{1/2}}{m^{1/r_1}} \rightarrow \infty,\ 
           \dfrac{\beta p}{n_0^{c_3}} \rightarrow 0
       \right\} \quad (r_1 \in \bN),
\end{align}
where $c_3$ is the constant given in Assumption A3.
A $GC_p$ criterion with $\alpha$ in $\cA$ is referred to as a high-dimensionality-adjusted consistent $GC_p$ ($HCGC_p$) criterion. (This name is based on, e.g.,  \citealp{Yanagihara2016} and \citealp{OdaYanagihara2020}.)
We have the following theorem on the JTT method based on the $HCGC_p$ criterion (the proof is given in Appendix~\ref{ap main}).
\begin{Thm} {\it 
\label{th main}
  Suppose that Assumptions A1--A3 hold.
  Then, under the asymptotic framework in \eqref{AFW}, we have 
  \begin{align*}
    \forall \alpha \in \cA,\ 
    \Pr \left( \hat{E} (\alpha) = E_\ast \right)
      \rightarrow 1.
  \end{align*}
  Moreover, the convergence order is given by
  \begin{align*}
    \Pr \left( \hat{E} (\alpha) = E_\ast \right) = 1 - \begin{dcases}
        O \left( m^2 \beta^{-2 r_1} p^{-r_1} + n_0^{2 - c_3 r_2} + m^{2 - r_2} n_0^{-r_2} \right)
          & (c_3 \ge 1) \\
        O \left( m^2 \beta^{-2 r_1} p^{-r_1} + n_0^{2 + r_2 - 2 c_3 r_2} \right)
          & (1/2 \le c_3 < 1) 
      \end{dcases},
  \end{align*}
  where $r_2$ is a natural number satisfying $r_2 > 2$.
}\end{Thm}
Theorem~\ref{th main} guarantees that the JTT method based on the $HCGC_p$ criterion has consistency in the edge selection.
To use this method in practice, we need to determine a specific value of $\alpha\ (\in \cA)$.
For example, $\alpha = 2$ expressing the $C_p$ criterion and $\alpha = \log n$ corresponding to the Bayesian information criterion (\citealp{Schwarz1978}) are often used, but they do not belong to $\cA$ (e.g., $\alpha = 2$ belongs to $\cA$ when $m$ is fixed and $p = O (\log n_0)$, and $\alpha = \log n$ belongs to $\cA$ when $n = m n_0$ and $p$ and $m$ are fixed).
As a specific $\alpha$ value satisfying the conditions of $\cA$, we propose 
\begin{align}
\label{alpha}
  \hat{\alpha}
    = \dfrac{N}{N - 2} + \hat{\beta}, \quad
  \hat{\beta}
    = B \cdot \dfrac{m^{1/4} \log n_0}{\sqrt{p}},\quad 
  B = \dfrac{N \sqrt{N + p - 2}}{(N - 2) \sqrt{N - 4}}.
\end{align}
If $r_1 \ge 4$ and $c_3 > 3/4$, then $\hat{\alpha} \in \cA$.
Hence, the probability that $\hat{E} (\hat{\alpha})$ is equal to $E_\ast$ converges to 1.
Notice that $B$ in $\hat{\beta}$ does not affect the consistency result because $B$ is of constant order.
Based on \cite{Yanagihara2016}, $B$ is incorporated to standardize $HCGC_p^{(k \ell)} (\hat{\alpha}) - HCGC_p^{(0)} (\hat{\alpha})$ for $(k, \ell) \in E_\ast$ (the details are given in Appendix~\ref{ap B}).

We can obtain another condition for the consistency of the JTT method based on a $GC_p$ criterion as  the following theorem (the proof is given in Appendix~\ref{ap sub1}).
\begin{Thm} {\it
\label{th sub1}
  Suppose that Assumptions A1--A3 hold.
  Furthermore, we define a class of $\alpha$ in $GC_p$ criteria as
  \begin{align*}
    \check{\cA}
      = \left\{
           \alpha > \dfrac{2}{1 - r}\ \middle|\ 
           \dfrac{\alpha p}{\log m} \rightarrow \infty,\ 
           \dfrac{\alpha p}{n_0^{c_3}} \rightarrow 0
         \right\} \quad (r \in (0, 1)).
  \end{align*}
  Then, under the asymptotic framework in \eqref{AFW}, we have 
  \begin{align*}
    \forall \alpha \in \check{\cA},\ 
    \Pr \left( \hat{E} (\alpha) = E_\ast \right)
      \rightarrow 1.
  \end{align*}
  Moreover, the convergence order is given by
  \begin{align*}
    &\Pr \left( \hat{E} (\alpha) = E_\ast \right) \\
    &\quad = 1 - O \left(
           \exp \left[
             - \alpha h p \left\{ (1 - r_1) - 1/\alpha \right\}
           \right] + m^2 \left\{ 
             \exp (- r_1 N / 4) + \exp (- c_1 c_2 n_0^{c_3} / 8) + \exp (- h r_2 N)
           \right\} 
         \right),
  \end{align*}
  where $h = (1 - \log 2) / 2$, $r_1 \in (0, 1)$, and $r_2 \in [1, \infty)$.
}\end{Thm}
Although the class of $\alpha$ given in Theorem~\ref{th sub1} looks like a relaxed version of $\cA$ in \eqref{class}, it is difficult to determine a specific $\alpha$ value due to the condition $\alpha > 2 / (1 - r)$.
For example, consider $\check{\alpha}$ given by
\begin{align*}
  \check{\alpha}
    = 2 + \dfrac{m^{1/4} \log n_0}{\sqrt{p}}.
\end{align*}
If $p / n_0^{c_3} \to 0$, then the probability that $\hat{E} (\check{\alpha})$ is equal to $E_\ast$ converges to 1.

We have already discussed a condition that guarantees consistency of the JTT method. Here, we will discuss a condition under which it is not consistent.
The condition for inconsistency of the JTT method is given by the following theorem (the proof is given in Appendix~\ref{ap sub2}).
\begin{Thm} {\it
\label{th sub2}
  Suppose that Assumptions A1--A3 hold.
  Then given the two conditions
  \begin{align*}
    {\rm C1}: \begin{dcases}
        \alpha \not\to \infty & (\text{$p$: fixed}) \\
        \lim \inf \alpha < 1 & (p \to \infty)
      \end{dcases},\quad
    {\rm C2}: \begin{dcases}
        \alpha p / \delta_{\min} \to \infty & (\delta_{\min} / p \to \infty) \\
        \lim \sup \alpha > 1 + c_4 & (\delta_{\min} / p \to c_4 \in [0, \infty))
      \end{dcases},
  \end{align*}
  under the asymptotic framework in \eqref{AFW}, for $\alpha$ satisfying either C1 or C2, we have
  \begin{align*}
    \Pr \left( \hat{E} (\alpha) = E_\ast \right)
      \not\to 1.
  \end{align*}
}\end{Thm}
For example, $\alpha = 2$ guarantees inconsistency when $p$ is fixed or there exists $c_4 \in [0, 1)$ (e.g., $\delta_{\min} = d_1 n_0$ and $p = (d_1 + \epsilon) n_0$ for $d_1 \in (0, 1)$ and $\epsilon \in (0, 1 - d_1)$), and $\alpha = \log n$ guarantees inconsistency when there exists $c_4 \in [0, \infty)$ (e.g., $\delta_{\min} = d_2 n_0$ and $p = d_3 n_0$ for $d_2 > 0$ and $d_3 \in (0, 1)$).

\subsection{Post-selection estimation}
\label{sec PSE}

In this section, we discuss an estimation method for $\vbeta_1, \ldots, \vbeta_m$ after obtaining the optimal edge set $\hat{E} (\alpha)$ selected by the JTT method.
One of simplest methods is the cluster-wise OLS method.
Let $\hat{m}$ be the number of connected components of the graph $(V, \hat{E} (\alpha))$ and $\hat{V}_i\ (i \in \{ 1, \ldots, \hat{m} \})$ be a vertex set of the $i$th connected component, where $\hat{V}_1, \ldots, \hat{V}_{\hat{m}}$ are nonempty and mutually exclusive sets satisfying $V = \cup_{i=1}^{\hat{m}} \hat{V}_i$.
This notation means that the JTT method produced  $\hat{m}$ clusters and $\hat{V}_i$ expresses the $i$th cluster.
Let $\tilde{\vy}_i$ and $\tilde{\vX}_i\ (i \in \{ 1, \ldots, \hat{m} \})$ be an $\tilde{n}_i$-dimensional vector and $\tilde{n}_i \times p$ matrix constructed by vertically stacking, respectively, the vectors $\vy_j$ and matrices $\vX_j$ for $j \in \hat{V}_i$, where $\tilde{n}_i = \sum_{j \in \hat{V}_i} n_j$.
Then, the cluster-wise OLS estimator of the regression coefficient vector for the $i$th cluster is given by
\begin{align*}
  \hat{\vxi}_i = \left( \tilde{\vX}_i' \tilde{\vX}_i \right)^{-1} \tilde{\vX}_i' \tilde{\vy}_i\quad (i \in \{ 1, \ldots, \hat{m} \}).
\end{align*}
If $\hat{E} (\alpha) = E_\ast$, then $\hat{\vxi}_i$ is a good estimator of $\vxi_i^\ast$.
However, its estimation accuracy may be low when the sample size within the cluster is small.

We try to improve the estimation accuracy by estimating with information of other clusters.
Specifically, in order to maintain the simplicity of the JTT method, a method with low calculation cost is desirable.
To achieve this, we employ a penalized estimation method with a weighted average of connected clusters.
Let $\hat{F}$ be an edge set for the clusters based on the original edge set $E$. 
Specifically, $\hat{F}$ is the edge set for the vertex set $\{ 1, \ldots, \hat{m} \}$ and $(k, \ell) \in \hat{F} \Rightarrow k < \ell$.
Given $\hat{F}$, we define a weighted average as
\begin{align*}
  \vb_i 
    = \dfrac{\sum_{(i, \ell) \in \hat{F}} w_{i \ell} \hat{\vxi}_\ell + \sum_{(k, i) \in \hat{F}} w_{i k} \hat{\vxi}_k}{
         \sum_{(i, \ell) \in \hat{F}} w_{i \ell} + \sum_{(k, i) \in \hat{F}} w_{i k}
       }, \quad
  w_{k \ell}
    = \left\| \hat{\vxi}_k - \hat{\vxi}_\ell \right\|^{-1}.
\end{align*}
Then, the estimator for the $i$th cluster is defined by
\begin{align*}
  \hat{\vxi}_i (\lambda)
    = \arg \min_{\vxi_i} \left\{
         \| \tilde{\vy}_i - \tilde{\vX}_i \vxi_i \|^2 + \lambda \| \vxi_i - \vb_i \|^2
       \right\}
    = \left( \tilde{\vX}_i' \tilde{\vX}_i + \lambda \vI_p \right)^{-1} \left( \tilde{\vX}_i' \tilde{\vy}_i + \lambda \vb_i \right),
\end{align*}
where $\lambda > 0$ is a tuning parameter.
By selecting $\lambda$ appropriately, we can expect to improve the estimation accuracy.
Here, we consider selecting $\lambda$ based on prediction accuracy, i.e., predictive mean square error (PMSE).
Let $\widehat{\tilde{\vy}}_i (\lambda)$ be a vector of fitted values obtained from $\hat{\vxi}_i (\lambda)$, i.e., $\widehat{\tilde{\vy}}_i (\lambda) = \tilde{\vX}_i \hat{\vxi}_i (\lambda)$.
Based on \cite{FujikoshiSatoh1997}, the modified $C_p$ ($MC_p$) criterion is given by
\begin{align*}
  MC_p (\lambda \mid i)
    &= \dfrac{1}{s_i^2} \| \tilde{\vy}_i - \widehat{\tilde{\vy}}_i (\lambda) \|^2
        + \dfrac{2 (\tilde{n}_i - p)}{\tilde{n}_i - p - 2} \tr \vH_i (\lambda), \\
  s_i^2
    &= \dfrac{\| \tilde{\vy}_i - \widehat{\tilde{\vy}}_i (0) \|^2}{\tilde{n}_i - p}, \quad
  \vH_i (\lambda)
    = \left( \tilde{\vX}_i' \tilde{\vX}_i + \lambda \vI_p \right)^{-1} \tilde{\vX}_i' \tilde{\vX}_i,
\end{align*}
and we define the optimal tuning parameter for the $i$th cluster as
\begin{align*}
  \hat{\lambda}_i
    = \arg \min_\lambda MC_p (\lambda \mid i).
\end{align*}
Notice that $MC_p (\lambda \mid i)$ is an unbiased estimator of PMSE for $\widehat{\tilde{\vy}}_i (\lambda)$ by regarding $w_{i \ell}$ as a constant.
Hence, we can expect to improve the prediction accuracy by selecting $\lambda$ minimizing $MC_p (\lambda \mid i)$  (actually, since $w_{i \ell}$ depends on $\tilde{\vy}_i$, $MC_p (\lambda \mid i)$ is a naive estimator).
On the other hand, because obtaining $\hat{\lambda_i}$ requires a numerical search and the inverse matrix in $MC_p (\lambda \mid i)$ depends on $\lambda$, this estimation method seems not to have a low calculation cost.
Fortunately, we can avoid calculating the inverse matrix by using a singular value decomposition of $\tilde{\vX}_i$.
Note that $\tilde{\vX}_i$ can be decomposed as
\begin{align*}
  \tilde{\vX}_i
    = \vU_i \vD_i^{1/2} \vV_i', \quad
  \vD_i
    = \diag (d_{i 1}, \ldots, d_{i p}),
\end{align*}
where $\vU_i$ is an $\tilde{n}_i \times p$ matrix, $\vV_i$ is an orthogonal matrix of order $p$, and the diagonal elements of $\vD_i$ satisfy $d_{i 1} \ge \cdots \ge d_{i p} > 0$.
Then, the $MC_p$ criterion can be rewritten as follows (e.g., \citealp{Yanagihara2018}):
\begin{align*}
  MC_p (\lambda \mid i)
    = \tilde{n}_i - p + \dfrac{1}{s_i^2} \sum_{j=1}^p \left( \dfrac{\lambda}{d_{i j} + \lambda} \right)^2 (z_{i j} - d_{i j} r_{i j})^2
         + \dfrac{2 (\tilde{n}_i - p)}{\tilde{n}_i - p - 2} \left( p - \sum_{j=1}^p \dfrac{\lambda}{d_{i j} + \lambda} \right),
\end{align*}
where $z_{i j}$ and $r_{i j}$ are the $j$th elements of $\vU_i' \tilde{\vy}_i$ and $\vD_i^{-1/2} \vV_i' \vb_i$, respectively.
Since this expression eliminates the inverse matrix, searching for $\hat{\lambda}_i$ numerical is simpler.

\section{Simulation}

In this section, we evaluate the performance of the JTT method proposed in this paper through Monte Carlo simulation with 1000 iterations. 
We also compare it with network Lasso.
The numerical calculation programs are executed in R (ver.~4.5.0; \citealp{R450}) on a computer running the Windows 11 Pro operating system with an AMD EPYC TM 7763 processor and 128 GB of RAM.
Network Lasso is implemented as R package \texttt{GGFL} (ver.~1.0.2; \citealp{GGFL102}), which employs BCDM (\citealp{Ohishi-2025}).
The JTT method is available via R package \texttt{JTT} (ver.~0.1.0; \citealp{JTT010}).
Note that although the \texttt{JTT} package includes some C++ code, the numerical calculation is conducted without the C++ code for a fair comparison of runtime (the \texttt{GGFL} package consists only of R code).

We first describe the setting of the simulation model.
Let $(V, E)$, which expresses the relationships among the $m$ groups, be a complete graph, and define sets $V_i^\ast\ (i \in \{ 1, \ldots, m_\ast \})$, which express the true clusters, for $m_\ast / m \in \{ 0.3, 0.6 \}\ (m \in \{ 20, 50 \})$, where $V = \{ 1, \ldots, m \}$.
Although the definition of each $V_i^\ast$ is omitted, they are available in the \texttt{JTT} package and the definition for $m = 20$ is the same as  that in \cite{Ohishi-2021}.
Then, the simulation model is defined by
\begin{align*}
  \vy_j \sim N_{n_0} (\vX_j \vbeta_j^\ast, \vI_{n_0})\ (j \in V), \quad
  \vbeta_j^\ast = \vxi_i^\ast = \nu i \1_p\ (j \in V_i^\ast;\ i \in \{ 1, \ldots, m_\ast \}),
\end{align*}
where $\vX_j = (\1_{n_0}, \vZ_j \vPsi (0.5)^{1/2})$, $\vZ_j$ is an $n_0 \times (p - 1)$ matrix with elements identically and independently distributed according to $U (-1, 1)$, $\vPsi (\rho)$ is a matrix of order $p - 1$ with $(i, j)$th elements $\rho^{|i-j|}$, $\nu$ is a constant adjusting the signal-to-noise ratio (SNR), and $\1_p$ is a $p$-dimensional vector of ones.
Here, SNR is defined by
\begin{align*}
  \mathrm{SNR}
    = \dfrac{1}{\# (F_\ast)} \sum_{(k, \ell) \in F_\ast} \dfrac{\Var[\vx' (\vxi_k^\ast - \vxi_\ell^\ast)]}{(p-1) \sigma_\ast^2}
    = \dfrac{1}{3 (p-1) \# (F_\ast)} \sum_{(k, \ell) \in F_\ast} (\vtheta_k^\ast - \vtheta_\ell^\ast)' \vPsi (0.5) (\vtheta_k^\ast - \vtheta_\ell^\ast),
\end{align*}
where $F_\ast$ is the edge set for the vertex set $\{ 1, \ldots, m_\ast \}$ of the true clusters, $\vx = (1, \vPsi (0.5)^{1/2} \vz')'$, $\vz$ is a $(p - 1)$-dimensional vector with elements identically and independently distributed according to $U (-1, 1)$, $\sigma_\ast^2 = 1$, and $\vtheta_k^\ast$ is the vector obtained by removing the first element from $\vxi_k^\ast$.
As described in Assumption A3, the difference of the true regression coefficient vectors for two groups is important for guaranteeing the consistency of the JTT method.
Hence, SNR is defined based on this difference.
Furthermore, to slow down the increase in SNR as $p$ increases, SNR is standardized by $p - 1$.
In this simulation, $\nu$ is defined to satisfy $\mathrm{SNR} = 3$.

Under the setting described above, we evaluate clustering accuracy, mean square error (MSE), and runtime.
Clustering accuracy here means the rate (\%) at which the true clusters are selected over 1000 iterations.
MSEs are defined for a vector of fitted values $\vy^\dagger = ({\vy_1^\dagger}', \ldots, {\vy_m^\dagger}')'$ and an estimator of regression coefficient vector $\vbeta^\dagger = ({\vbeta_1^\dagger}', \ldots, {\vbeta_m^\dagger}')'$, respectively, as
\begin{align*}
  \MSE_{\rm f} [\vy^\dagger] 
    &= \dfrac{1}{n} \E \left[ \sum_{j=1}^m \| \vX_j \vbeta_j^\ast - \vy_j^\dagger \|^2 \right], \quad
  \MSE_{\rm c} [\vbeta^\dagger]
    = \dfrac{1}{mp}\E \left[ \sum_{j=1}^m \| \vbeta_j^\ast - \vbeta_j^\dagger \|^2 \right],
\end{align*}
where expectation is evaluated by Monte Carlo simulation with 1000 iterations.
In the simulation, MSEs for $\vy_j^\dagger = \vX_j \vbeta_j^\dagger$ and $\vbeta_j^\dagger = \hat{\vbeta}_j$ are given by $\MSE_{\rm f} [\vy^\dagger] = mp/n = p / n_0$ and $\MSE_{\rm c} [\vbeta^\dagger] = \sum_{j=1}^m \tr (\vM_j^{-1}) / mp$, respectively, where $\hat{\vbeta}_j$ is the group-wise OLS estimator of $\vbeta_j$ given in \eqref{gwLSE}.
The methods used in this simulation are the JTT method and network Lasso.
The $HCGC_p$ criterion with $\hat{\alpha}$ in \eqref{alpha} is used in the JTT method.
Furthermore, the cluster-wise OLS method and the penalized estimation method described in Section~\ref{sec PSE} are applied as an estimation method after clustering, denoted as JTT1 and JTT2, respectively.
Network Lasso is implemented with the \texttt{GGFL} package under default settings.
The penalty weights are thus  given by $w_{k \ell} = \| \hat{\vbeta}_k - \hat{\vbeta}_\ell \|^{-1}$, and tuning parameter $\lambda$ is selected based on minimizing the following extended GCV (EGCV) criterion (\citealp{Ohishi-2020}):
\begin{align*}
  \mathrm{EGCV} (\lambda \mid \alpha)
    = \dfrac{\sum_{j=1}^m \| \vy_j - \hat{\vy}_{\lambda, j} \|^2}{( 1 - \df_\lambda / n )^\alpha},
\end{align*}
where $\hat{\vy}_{\lambda, j}$ is the vector of fitted values obtained from the network Lasso estimator, $\df_\lambda$ is the number of unique parameters, and $\alpha = \log n$.
We denote network Lasso as NL in the presented results.

\begin{table}[h]
\caption{\ Clustering accuracy (\%) for fixed $p$ \label{tab CA1}}
\centering
\begin{tabular}{rrrrrrrrrr}
\toprule
  &  & \multicolumn{4}{c}{$p = 20$} & \multicolumn{4}{c}{$p = 40$} \\
\cmidrule(lr){3-6} \cmidrule(lr){7-10}
  &  & \multicolumn{2}{c}{$m_\ast / m = 0.3$} & \multicolumn{2}{c}{$m_\ast / m = 0.6$} & \multicolumn{2}{c}{$m_\ast / m = 0.3$} & \multicolumn{2}{c}{$m_\ast / m = 0.6$} \\
\cmidrule(lr){3-4} \cmidrule(lr){5-6} \cmidrule(lr){7-8} \cmidrule(lr){9-10}
 $m$ & $n_0$ & JTT & NL & JTT & NL & JTT & NL & JTT & NL \\
\midrule
 20 & 50 & \bf 100.0 & 0.0 & 0.0 & 0.0 & \bf 100.0 & 0.0 & \bf 0.3 & 0.0 \\
  & 100 & \bf 100.0 & 30.3 & \bf 95.0 & 10.8 & \bf 100.0 & 2.6 & \bf 100.0 & 1.4 \\
  & 200 & \bf 100.0 & 76.9 & \bf 100.0 & 35.6 & \bf 100.0 & 71.8 & \bf 100.0 & 45.6 \\
  & 500 & \bf 100.0 & 93.6 & \bf 100.0 & 80.5 & \bf 100.0 & 97.9 & \bf 100.0 & 94.0 \\
  & 1000 & \bf 100.0 & 99.2 & \bf 100.0 & 98.0 & \bf 100.0 & \bf 100.0 & \bf 100.0 & 99.1 \\ \cmidrule{1-10}
 50 & 50 & 0.0 & 0.0 & 0.0 & 0.0 & 0.0 & 0.0 & 0.0 & 0.0 \\
  & 100 & 0.0 & 0.0 & 0.0 & 0.0 & \bf 1.1 & 0.0 & 0.0 & 0.0 \\
  & 200 & \bf 97.7 & 0.1 & 0.0 & 0.0 & \bf 100.0 & 0.0 & 0.0 & 0.0 \\
  & 500 & \bf 100.0 & 8.7 & 0.0 & \bf 0.6 & \bf 100.0 & 9.1 & \bf 89.7 & 2.5 \\
  & 1000 & \bf 100.0 & 23.0 & \bf 100.0 & 8.2 & \bf 100.0 & 30.6 & \bf 100.0 & 18.2 \\
\bottomrule
\end{tabular}
\end{table}
\begin{table}[h]
\caption{\ Clustering accuracy (\%)  for fixed $p/n_0$ \label{tab CA2}}
\centering
\begin{tabular}{rrrrrrrrrr}
\toprule
  &  & \multicolumn{4}{c}{$p/n_0 = 0.4$} & \multicolumn{4}{c}{$p/n_0 = 0.8$} \\
\cmidrule(lr){3-6} \cmidrule(lr){7-10}
  &  & \multicolumn{2}{c}{$m_\ast / m = 0.3$} & \multicolumn{2}{c}{$m_\ast / m = 0.6$} & \multicolumn{2}{c}{$m_\ast / m = 0.3$} & \multicolumn{2}{c}{$m_\ast / m = 0.6$} \\
\cmidrule(lr){3-4} \cmidrule(lr){5-6} \cmidrule(lr){7-8} \cmidrule(lr){9-10}
 $m$ & $n_0$ & JTT & NL & JTT & NL & JTT & NL & JTT & NL \\
\midrule
 20 & 50 & \bf 100.0 & 0.6 & 0.0 & 0.0 & \bf 100.0 & 0.0 & \bf 1.3 & 0.0 \\
  & 100 & \bf 100.0 & 1.7 & \bf 100.0 & 1.3 & \bf 100.0 & 0.0 & \bf 99.9 & 0.0 \\
  & 200 & \bf 100.0 & 14.2 & \bf 100.0 & 22.5 & \bf 100.0 & 0.0 & \bf 100.0 & 0.0 \\
  & 500 & \bf 100.0 & 72.8 & \bf 100.0 & 92.8 & \bf 100.0 & 0.0 & \bf 100.0 & 0.0 \\
  & 1000 & \bf 100.0 & \bf 100.0 & \bf 100.0 & \bf 100.0 & \bf 100.0 & 4.1 & \bf 100.0 & 20.3 \\ \cmidrule{1-10}
 50 & 50 & 0.0 & 0.0 & 0.0 & 0.0 & 0.0 & 0.0 & 0.0 & 0.0 \\
  & 100 & \bf 1.4 & 0.0 & 0.0 & 0.0 & \bf 10.1 & 0.0 & 0.0 & 0.0 \\
  & 200 & \bf 100.0 & 0.0 & 0.0 & 0.0 & \bf 100.0 & 0.0 & 0.0 & 0.0 \\
  & 500 & \bf 100.0 & 0.0 & \bf 100.0 & 0.0 & \bf 100.0 & 0.0 & \bf 100.0 & 0.0 \\
  & 1000 & \bf 100.0 & 7.8 & \bf 100.0 & 0.0 & \bf 100.0 & 0.0 & \bf 100.0 & 0.0 \\
\bottomrule
\end{tabular}
\end{table}
Tables~\ref{tab CA1} and \ref{tab CA2} summarize clustering accuracy for fixed $p$ and $p/n_0$, respectively. In these tables, bold font indicates the higher clustering accuracy.
As shown in Table~\ref{tab CA1}, for fixed $p$, the JTT method performed better than network Lasso in clustering, with its accuracy reaching 100\% at a much smaller $n_0$.
For $m = 50$ and $m_\ast / m = 0.6$, neither method could select the true clusters for smaller $n_0$.
Actually, $\min_{(k, \ell) \notin E_\ast} \| \vxi_k^\ast - \vxi_\ell^\ast \|^2 / \sigma_\ast^2$ is extremely small, making the determination of the true clusters very difficult.
In this situation, the JTT method unified all groups into one cluster in most cases.
Network Lasso behaved differently.
For example, for $m = 50$, $m_\ast / m = 0.6$, and $n_0 = 100, 200$, network Lasso determined that every group defines its own cluster.
Although network Lasso could not determine the true clusters like the JTT method, it performed a conservative selection by selecting the smallest $\lambda$.
However, the JTT method was able to select the true clusters for sufficiently large $n_0$.
One the other hand, when $p$ increases with $n_0$, as shown in Table~\ref{tab CA2}, the JTT method maintained high clustering accuracy, whereas network Lasso was not able to select the true clusters, particularly for large $p$ or $m$.

\begin{table}[h]
\caption{\ Relative MSE (\%) for $p = 20$ \label{tab MSE1-20}}
\centering
{\setlength{\tabcolsep}{4.1pt}
\begin{tabular}{rrrrrrrrrrrrrr}
\toprule
  &  & \multicolumn{6}{c}{$m_\ast / m = 0.3$} & \multicolumn{6}{c}{$m_\ast / m = 0.6$} \\
\cmidrule(lr){3-8} \cmidrule(lr){9-14}
  &  & \multicolumn{3}{c}{$\RMSE_{\rm f}$} & \multicolumn{3}{c}{$\RMSE_{\rm c}$} & \multicolumn{3}{c}{$\RMSE_{\rm f}$} & \multicolumn{3}{c}{$\RMSE_{\rm c}$} \\
\cmidrule(lr){3-5} \cmidrule(lr){6-8} \cmidrule(lr){9-11} \cmidrule(lr){12-14}
 $m$ & $n_0$ & JTT1 & JTT2 & NL & JTT1 & JTT2 & NL & JTT1 & JTT2 & NL & JTT1 & JTT2 & NL \\
\midrule
 20 & 50 & 30.0 & \bf 24.6 & 286.7 & 20.6 & \bf 13.8 & 58.6 & 921.0 & 921.7 & 104.0 & 176.2 & 162.4 & \bf 98.7 \\
  & 100 & 30.3 & \bf 26.1 & 144.4 & 25.7 & \bf 19.1 & 37.3 & 62.0 & \bf 46.9 & 159.7 & 55.7 & \bf 31.9 & 58.8 \\
  & 200 & 30.9 & \bf 28.0 & 87.9 & 28.3 & \bf 23.0 & 30.2 & 60.6 & \bf 48.6 & 129.8 & 58.0 & \bf 38.1 & 42.0 \\
  & 500 & 32.4 & \bf 30.7 & 56.4 & 29.8 & \bf 26.4 & 27.3 & 62.3 & \bf 53.9 & 93.9 & 59.7 & 45.1 & \bf 39.7 \\
  & 1000 & 34.9 & \bf 33.8 & 46.8 & 30.7 & 28.5 & \bf 27.9 & 64.5 & \bf 58.3 & 78.3 & 60.3 & 49.1 & \bf 41.9 \\ \cmidrule{1-14}
 50 & 50 & 6159.5 & 6159.5 & 3708.9 & 857.4 & 857.4 & 542.7 & 6600.9 & 6600.9 & 3561.4 & 866.7 & 866.6 & 507.8 \\
  & 100 & 2259.5 & 2260.5 & 146.0 & 429.9 & 425.0 & \bf 97.4 & 12752.5 & 12752.5 & 100.1 & 2324.6 & 2324.6 & \bf 99.6 \\
  & 200 & 31.0 & \bf 25.6 & 287.1 & 28.0 & \bf 19.7 & 69.8 & 21107.2 & 21107.5 & 100.7 & 4314.9 & 4314.4 & \bf 99.9 \\
  & 500 & 29.9 & \bf 26.0 & 166.1 & 29.3 & \bf 23.0 & 45.8 & 297.5 & 290.0 & 180.6 & 101.0 & 87.7 & \bf 58.0 \\
  & 1000 & 29.9 & \bf 26.8 & 110.7 & 29.5 & \bf 24.7 & 35.6 & 65.1 & \bf 56.4 & 156.8 & 60.7 & \bf 45.7 & 47.7 \\
\bottomrule
\end{tabular}}
\end{table}
\begin{table}[h]
\caption{\ Relative MSE (\%) for $p = 40$ \label{tab MSE1-40}}
\centering
{\setlength{\tabcolsep}{4.1pt}
\begin{tabular}{rrrrrrrrrrrrrr}
\toprule
  &  & \multicolumn{6}{c}{$m_\ast / m = 0.3$} & \multicolumn{6}{c}{$m_\ast / m = 0.6$} \\
\cmidrule(lr){3-8} \cmidrule(lr){9-14}
  &  & \multicolumn{3}{c}{$\RMSE_{\rm f}$} & \multicolumn{3}{c}{$\RMSE_{\rm c}$} & \multicolumn{3}{c}{$\RMSE_{\rm f}$} & \multicolumn{3}{c}{$\RMSE_{\rm c}$} \\
\cmidrule(lr){3-5} \cmidrule(lr){6-8} \cmidrule(lr){9-11} \cmidrule(lr){12-14}
 $m$ & $n_0$ & JTT1 & JTT2 & NL & JTT1 & JTT2 & NL & JTT1 & JTT2 & NL & JTT1 & JTT2 & NL \\
\midrule
 20 & 50 & 29.9 & \bf 23.6 & 4919.4 & 7.7 & \bf 4.7 & 246.0 & 1132.0 & 1142.8 & 5248.8 & 77.3 & \bf 63.9 & 266.5 \\
  & 100 & 30.0 & \bf 25.4 & 212.6 & 20.7 & \bf 14.8 & 41.2 & 59.8 & \bf 43.3 & 211.1 & 50.4 & \bf 26.0 & 67.2 \\
  & 200 & 30.3 & \bf 27.1 & 102.6 & 25.8 & \bf 20.6 & 29.4 & 60.2 & \bf 47.1 & 149.2 & 55.9 & \bf 35.0 & 42.4 \\
  & 500 & 31.2 & \bf 29.4 & 56.3 & 28.6 & \bf 25.0 & 25.7 & 61.1 & \bf 51.8 & 89.8 & 58.6 & 42.9 & \bf 36.4 \\
  & 1000 & 32.4 & \bf 31.3 & 43.5 & 29.6 & 27.2 & \bf 26.3 & 62.2 & \bf 55.4 & 70.1 & 59.4 & 47.5 & \bf 39.1 \\ \cmidrule{1-14}
 50 & 50 & 5843.6 & 5844.1 & 6041.4 & 248.1 & 247.9 & 256.5 & 6018.1 & 6018.1 & 6018.1 & 264.0 & 264.0 & 264.0 \\
  & 100 & 169.3 & 164.8 & 199.6 & 38.7 & \bf 31.2 & 101.6 & 12353.1 & 12353.2 & 112.4 & 1649.7 & 1649.5 & 101.6 \\
  & 200 & 30.0 & \bf 24.0 & 343.9 & 26.0 & \bf 17.6 & 70.8 & 11738.4 & 11743.1 & 100.2 & 2101.1 & 2095.6 & \bf 99.7 \\
  & 500 & 30.0 & \bf 25.7 & 188.7 & 28.4 & \bf 21.9 & 45.7 & 62.8 & \bf 50.4 & 225.2 & 59.0 & \bf 38.8 & 63.0 \\
  & 1000 & 30.0 & \bf 26.6 & 115.6 & 29.2 & \bf 24.1 & 33.3 & 62.5 & \bf 52.9 & 161.6 & 59.9 & \bf 43.7 & 44.6 \\
\bottomrule
\end{tabular}}
\end{table}
Tables~\ref{tab MSE1-20}--\ref{tab MSE2-08} summarize MSE for both the fitted values and estimator of regression coefficients of which values are relative MSE (RMSE; \%) based on MSE obtained from the group-wise OLS estimator
\begin{align*}
  \RMSE_{\rm f} [\vy^\dagger]
    &= 100 \times \dfrac{\MSE_{\rm f} [\vy^\dagger]}{mp/n}, \quad
  \RMSE_{\rm c} [\vbeta^\dagger]
    = 100 \times \dfrac{\MSE_{\rm c} [\vbeta^\dagger]}{\sum_{j=1}^m \tr (\vM_j^{-1}) / mp}.
\end{align*}
An RMSE value smaller than 100 means that the method is superior to the group-wise OLS method in estimation accuracy.
In these tables, bold font indicates the minimum value, if any value is less than 100.
Notice that under the true clusters, MSE for the fitted values of JTT1 is $m_\ast p / n$.
Hence, RMSE for the fitted values of JTT1 is $m_\ast / m$ if the clustering accuracy of the JTT method is 100\% (since the values in the tables are approximations from Monte Carlo simulation, they do not coincide exactly with $m_\ast / m$ even when the clustering accuracy is 100\%).
In the tables, we can see that JTT2 had the best estimation accuracy for both fitted values and regression coefficients for most cases.
Recall that JTT2 employs a simple penalized estimation method. 
Nevertheless, it improved MSE compared to JTT1 for most cases.
Although we can see that $\RMSE_{\rm f}$ of JTT1 when $p$ is fixed got worse as $n_0$ increased, this is a problem not of the estimation accuracy but the approximation accuracy of Monte Carlo simulation, because $\MSE_{\rm f}$ gets smaller as $n_0$ increases.
The same behavior was not observed for $p/n_0$ fixed, because $\MSE_{\rm f}$ remained constant as $n_0$ increased.
Moreover, when clustering accuracy was extremely low, i.e., it was difficult to determine the true clusters, the JTT method and network Lasso showed different tendencies.
For instance, for $m = 50$, $m_\ast / m = 0.6$, and $n_0 = 100, 200$ in Table~\ref{tab MSE1-40}, RMSE of the JTT method is extremely large, whereas that of network Lasso is close to 100.
This difference of RMSE can in turn be explained by the difference of clustering behaviors.
Since the JTT method unifies all groups into one cluster, MSE gets much worse (in this case, MSEs of JTT1 and JTT2 are equal).
In contrast, when network Lasso selects the smallest $\lambda$ and no groups are joined together, the estimator is close to the group-wise OLS estimator and so RMSE is close to 100.
\begin{table}[h]
\caption{\ Relative MSE (\%) for $p/n_0 = 0.4$ \label{tab MSE2-04}}
\centering
{\setlength{\tabcolsep}{4pt}
\begin{tabular}{rrrrrrrrrrrrrr}
\toprule
  &  & \multicolumn{6}{c}{$m_\ast / m = 0.3$} & \multicolumn{6}{c}{$m_\ast / m = 0.6$} \\
\cmidrule(lr){3-8} \cmidrule(lr){9-14}
  &  & \multicolumn{3}{c}{$\RMSE_{\rm f}$} & \multicolumn{3}{c}{$\RMSE_{\rm c}$} & \multicolumn{3}{c}{$\RMSE_{\rm f}$} & \multicolumn{3}{c}{$\RMSE_{\rm c}$} \\
\cmidrule(lr){3-5} \cmidrule(lr){6-8} \cmidrule(lr){9-11} \cmidrule(lr){12-14}
 $m$ & $n_0$ & JTT1 & JTT2 & NL & JTT1 & JTT2 & NL & JTT1 & JTT2 & NL & JTT1 & JTT2 & NL \\
\midrule
 20 & 50 & 30.1 & \bf 24.4 & 285.6 & 20.3 & \bf 13.3 & 49.0 & 536.0 & 532.9 & 118.0 & 110.4 & \bf 91.2 & 95.1 \\
  & 100 & 29.9 & \bf 25.3 & 216.7 & 20.5 & \bf 14.6 & 39.8 & 59.8 & \bf 43.2 & 199.4 & 50.7 & \bf 25.9 & 69.3 \\
  & 200 & 29.9 & \bf 26.4 & 151.7 & 20.5 & \bf 16.1 & 32.3 & 59.8 & \bf 45.5 & 207.2 & 50.6 & \bf 29.2 & 46.3 \\
  & 500 & 29.9 & \bf 27.8 & 84.8 & 20.6 & \bf 17.6 & 23.8 & 59.9 & \bf 49.2 & 122.0 & 50.9 & \bf 34.3 & 37.9 \\
  & 1000 & 30.0 & \bf 28.6 & 52.2 & 20.6 & \bf 18.7 & 20.3 & 59.9 & \bf 51.8 & 76.6 & 50.9 & 37.7 & \bf 33.9 \\ \cmidrule{1-14}
 50 & 50 & 5875.2 & 5875.3 & 5043.5 & 845.6 & 845.5 & 738.7 & 5846.1 & 5846.1 & 2241.9 & 867.8 & 867.8 & 384.7 \\
  & 100 & 220.2 & 216.2 & 265.6 & 45.5 & \bf 38.2 & 101.5 & 12518.2 & 12518.2 & \bf 99.9 & 1609.1 & 1609.1 & \bf 99.9 \\
  & 200 & 30.0 & \bf 23.7 & 571.7 & 21.3 & \bf 13.8 & 88.0 & 2906.6 & 2909.7 & \bf 99.9 & 407.6 & 395.2 & \bf 99.9 \\
  & 500 & 30.0 & \bf 25.2 & 295.3 & 21.3 & \bf 15.7 & 47.0 & 60.0 & \bf 46.0 & 259.1 & 51.6 & \bf 30.3 & 81.4 \\
  & 1000 & 30.0 & \bf 26.2 & 338.0 & 21.3 & \bf 16.9 & 55.8 & 60.0 & \bf 48.8 & 272.5 & 51.6 & \bf 34.0 & 52.6 \\
\bottomrule
\end{tabular}}
\end{table}
\begin{table}[h]
\caption{\ Relative MSE (\%) for $p/n_0 = 0.8$ \label{tab MSE2-08}}
\centering
{\setlength{\tabcolsep}{3.7pt}
\begin{tabular}{rrrrrrrrrrrrrr}
\toprule
  &  & \multicolumn{6}{c}{$m_\ast / m = 0.3$} & \multicolumn{6}{c}{$m_\ast / m = 0.6$} \\
\cmidrule(lr){3-8} \cmidrule(lr){9-14}
  &  & \multicolumn{3}{c}{$\RMSE_{\rm f}$} & \multicolumn{3}{c}{$\RMSE_{\rm c}$} & \multicolumn{3}{c}{$\RMSE_{\rm f}$} & \multicolumn{3}{c}{$\RMSE_{\rm c}$} \\
\cmidrule(lr){3-5} \cmidrule(lr){6-8} \cmidrule(lr){9-11} \cmidrule(lr){12-14}
 $m$ & $n_0$ & JTT1 & JTT2 & NL & JTT1 & JTT2 & NL & JTT1 & JTT2 & NL & JTT1 & JTT2 & NL \\
\midrule
 20 & 50 & 30.0 & \bf 23.8 & 3348.4 & 7.6 & \bf 4.8 & 164.1 & 586.9 & 586.6 & 5105.2 & 51.7 & \bf 38.9 & 242.7 \\
  & 100 & 30.0 & \bf 24.9 & 7885.0 & 7.9 & \bf 5.4 & 372.9 & 60.0 & \bf 42.6 & 11216.0 & 36.2 & \bf 11.9 & 546.0 \\
  & 200 & 29.9 & \bf 26.2 & 897.8 & 8.0 & \bf 6.1 & 65.2 & 60.0 & \bf 44.9 & 20216.5 & 36.1 & \bf 14.6 & 1028.5 \\
  & 500 & 30.0 & \bf 27.7 & 272.1 & 8.1 & \bf 6.8 & 25.5 & 60.0 & \bf 48.2 & 53670.8 & 36.2 & \bf 17.6 & 2481.2 \\
  & 1000 & 30.0 & \bf 28.5 & 144.6 & 8.1 & \bf 7.3 & 16.8 & 60.0 & \bf 50.7 & 73500.7 & 36.4 & \bf 21.1 & 3718.0 \\ \cmidrule{1-14}
 50 & 50 & 5989.4 & 5989.8 & 6053.9 & 245.9 & 245.7 & 245.7 & 5997.2 & 5997.2 & 5934.3 & 255.1 & 255.1 & 250.3 \\
  & 100 & 134.6 & 129.4 & 12369.6 & 14.6 & \bf 9.8 & 502.3 & 12436.0 & 12436.2 & 12531.6 & 530.7 & 530.6 & 534.9 \\
  & 200 & 30.0 & \bf 23.4 & 20950.9 & 9.4 & \bf 5.3 & 940.8 & 2652.1 & 2662.7 & 24446.2 & 139.1 & 127.2 & 1055.8 \\
  & 500 & 30.0 & \bf 25.0 & 57578.4 & 9.5 & \bf 6.2 & 2489.3 & 60.0 & \bf 45.2 & 61407.8 & 38.2 & \bf 15.5 & 2607.4 \\
  & 1000 & 30.0 & \bf 26.0 & 565.5 & 9.6 & \bf 6.8 & 43.2 & 60.0 & \bf 47.8 & 122132.5 & 38.1 & \bf 18.1 & 5155.1 \\
\bottomrule
\end{tabular}}
\end{table}

\begin{table}[h]
\caption{\ Rutime (sec.) for fixed $p$ \label{tab time1}}
\centering
\begin{tabular}{rrrrrrrrrrrrrr}
\toprule
  &  & \multicolumn{6}{c}{$p = 20$} & \multicolumn{6}{c}{$p = 40$} \\
\cmidrule(lr){3-8} \cmidrule(lr){9-14}
  &  & \multicolumn{3}{c}{$m_\ast / m = 0.3$} & \multicolumn{3}{c}{$m_\ast / m = 0.6$} & \multicolumn{3}{c}{$m_\ast / m = 0.3$} & \multicolumn{3}{c}{$m_\ast / m = 0.6$} \\
\cmidrule(lr){3-5} \cmidrule(lr){6-8} \cmidrule(lr){9-11} \cmidrule(lr){12-14}
 $m$ & $n_0$ & JTT1 & JTT2 & NL & JTT1 & JTT2 & NL & JTT1 & JTT2 & NL & JTT1 & JTT2 & NL \\
\midrule
 20 & 50 & \bf 0.1 & 0.1 & 21.0 & \bf 0.1 & 0.1 & 22.7 & \bf 0.1 & 0.2 & 31.8 & \bf 0.1 & 0.1 & 42.0 \\
  & 100 & \bf 0.1 & 0.1 & 16.1 & \bf 0.1 & 0.1 & 19.1 & \bf 0.1 & 0.1 & 21.4 & \bf 0.1 & 0.1 & 30.2 \\
  & 200 & \bf 0.1 & 0.1 & 14.1 & \bf 0.1 & 0.1 & 17.2 & \bf 0.1 & 0.2 & 16.6 & \bf 0.1 & 0.2 & 21.4 \\
  & 500 & \bf 0.1 & 0.1 & 12.4 & \bf 0.1 & 0.2 & 14.6 & \bf 0.1 & 0.3 & 14.3 & \bf 0.1 & 0.3 & 17.3 \\
  & 1000 & \bf 0.1 & 0.2 & 11.9 & \bf 0.1 & 0.3 & 13.8 & \bf 0.3 & 0.4 & 13.5 & \bf 0.2 & 0.5 & 16.2 \\ \cmidrule{1-14}
 50 & 50 & \bf 0.2 & 0.2 & 290.6 & \bf 0.2 & 0.2 & 344.2 & 0.3 & \bf 0.3 & 496.8 & 0.3 & \bf 0.3 & 532.8 \\
  & 100 & \bf 0.3 & 0.3 & 257.8 & 0.3 & \bf 0.3 & 317.8 & \bf 0.3 & 0.4 & 336.3 & 0.3 & \bf 0.3 & 418.6 \\
  & 200 & \bf 0.2 & 0.3 & 220.7 & \bf 0.3 & 0.3 & 288.2 & \bf 0.3 & 0.5 & 286.7 & \bf 0.3 & 0.4 & 369.6 \\
  & 500 & \bf 0.3 & 0.5 & 178.7 & \bf 0.3 & 0.5 & 252.7 & \bf 0.4 & 0.7 & 213.7 & \bf 0.5 & 0.8 & 317.0 \\
  & 1000 & \bf 0.4 & 0.7 & 158.1 & \bf 0.3 & 0.7 & 222.8 & \bf 0.6 & 1.1 & 183.3 & \bf 0.6 & 1.2 & 268.9 \\
\bottomrule
\end{tabular}
\end{table}
\begin{table}[h]
\caption{\ Runtime (sec.)  for fixed $p/n_0$ \label{tab time2}}
\centering
{\setlength{\tabcolsep}{5.1pt}
\begin{tabular}{rrrrrrrrrrrrrr}
\toprule
  &  & \multicolumn{6}{c}{$p/n_0 = 0.4$} & \multicolumn{6}{c}{$p/n_0 = 0.8$} \\
\cmidrule(lr){3-8} \cmidrule(lr){9-14}
  &  & \multicolumn{3}{c}{$m_\ast / m = 0.3$} & \multicolumn{3}{c}{$m_\ast / m = 0.6$} & \multicolumn{3}{c}{$m_\ast / m = 0.3$} & \multicolumn{3}{c}{$m_\ast / m = 0.6$} \\
\cmidrule(lr){3-5} \cmidrule(lr){6-8} \cmidrule(lr){9-11} \cmidrule(lr){12-14}
 $m$ & $n_0$ & JTT1 & JTT2 & NL & JTT1 & JTT2 & NL & JTT1 & JTT2 & NL & JTT1 & JTT2 & NL \\
\midrule
 20 & 50 & \bf 0.1 & 0.1 & 20.6 & \bf 0.1 & 0.1 & 23.8 & \bf 0.1 & 0.1 & 32.3 & \bf 0.1 & 0.1 & 34.6 \\
  & 100 & \bf 0.1 & 0.1 & 22.5 & \bf 0.1 & 0.1 & 28.1 & \bf 0.1 & 0.2 & 42.6 & \bf 0.1 & 0.2 & 91.8 \\
  & 200 & \bf 0.1 & 0.3 & 23.5 & \bf 0.1 & 0.3 & 32.2 & \bf 0.4 & 0.8 & 53.3 & \bf 0.4 & 0.9 & 79.4 \\
  & 500 & \bf 0.8 & 2.4 & 41.7 & \bf 0.8 & 2.6 & 55.9 & \bf 3.6 & 10.2 & 175.4 & \bf 3.6 & 10.9 & 309.2 \\
  & 1000 & \bf 5.0 & 19.1 & 110.1 & \bf 4.7 & 18.1 & 174.7 & \bf 26.0 & 99.6 & 726.1 & \bf 26.5 & 90.7 & 1976.2 \\ \cmidrule{1-14}
 50 & 50 & \bf 0.3 & 0.3 & 303.7 & \bf 0.2 & 0.2 & 335.6 & \bf 0.3 & 0.3 & 594.0 & \bf 0.3 & 0.3 & 641.8 \\
  & 100 & \bf 0.3 & 0.4 & 345.1 & 0.3 & \bf 0.3 & 399.2 & \bf 0.5 & 0.7 & 725.8 & \bf 0.5 & 0.6 & 1081.6 \\
  & 200 & \bf 0.7 & 1.0 & 426.5 & \bf 0.6 & 0.9 & 504.9 & \bf 1.7 & 3.0 & 950.6 & \bf 1.7 & 2.8 & 1205.5 \\
  & 500 & \bf 3.4 & 7.2 & 640.2 & \bf 3.5 & 8.1 & 828.7 & \bf 16.6 & 33.2 & 1782.7 & \bf 17.1 & 35.3 & 3602.5 \\
  & 1000 & \bf 19.6 & 54.9 & 1105.8 & \bf 20.5 & 60.4 & 1883.5 & \bf 128.3 & 465.5 & 6051.2 & \bf 147.0 & 519.6 & 10601.3 \\
\bottomrule
\end{tabular}}
\end{table}
Tables~\ref{tab time1} and \ref{tab time2} summarize runtime, with bold font indicating the smallest value.
As shown, JTT1 was much faster than network Lasso.
Furthermore, we can see that the penalized estimation in JTT2 does not affect runtime much.
Since the JTT method requires repeated calculation of the inverse of a $p \times p$ matrix, its runtime greatly increases as $m$ or $p$ increases.
However, this is not as major an issue as with network Lasso.
We have some instances of the impossible result  that JTT2 is faster than JTT1, but this result is caused by numerical error from runtime being extremely short.
Overall, the results of the simulation allow us to conclude that the JTT method has excellent performance compared to network Lasso in terms of clustering accuracy, estimation accuracy, and runtime.

\section{Real data example}
\label{sec realdata}

In this section, we compare the JTT method with network Lasso through an application to real data.
Settings and implementation environment are the same as in the previous section.
The dataset used is Ames Housing Data (\citealp{DeCock2011}) obtained from R package \texttt{modeldata} (ver.~1.5.1; \citealp{modeldata151}).
We regard a variable Neighborhood, which indicates a location of housing, as a group, and define the edge set $E$ by regional adjacency.
In this application, small sample groups are merged with other groups in advance.
Let the response variable be Sale\_Price (USD), which is housing price, and the remainder be explanatory variables.
Category variables among the explanatory variables are transformed to binary dummy variables with value 1 corresponding to the most frequent category.
Furthermore, some variables are removed from the model to guarantee full column rank of the matrix of explanatory variables for each group.
In the end, we apply the JTT method and network Lasso to Ames Housing Data with $n = 2930$, $p = 34$, $m = 19$, and $\# (E) = 31$, where the group-wise sample sizes are as summarized in Table~\ref{tab gwn}.
\begin{table}[h]
\caption{\ Group-wise sample sizes \label{tab gwn}}
\centering
{\setlength{\tabcolsep}{4pt}
\begin{tabular}{c|rrrrrrrrrrrrrrrrrrr}
\toprule
 Group & 1 & 2 & 3 & 4 & 5 & 6 & 7 & 8 & 9 & 10 & 11 & 12 & 13 & 14 & 15 & 16 & 17 & 18 & 19 \\
 Sample size & 496 & 268 & 239 & 194 & 214 & 166 & 193 & 151 & 131 & 125 & 151 & 108 & 113 & 93 & 74 & 71 & 51 & 48 & 44 \\
\bottomrule
\end{tabular}}
\end{table}

\begin{figure}[h]
  \centering
  \includegraphics[scale=0.33]{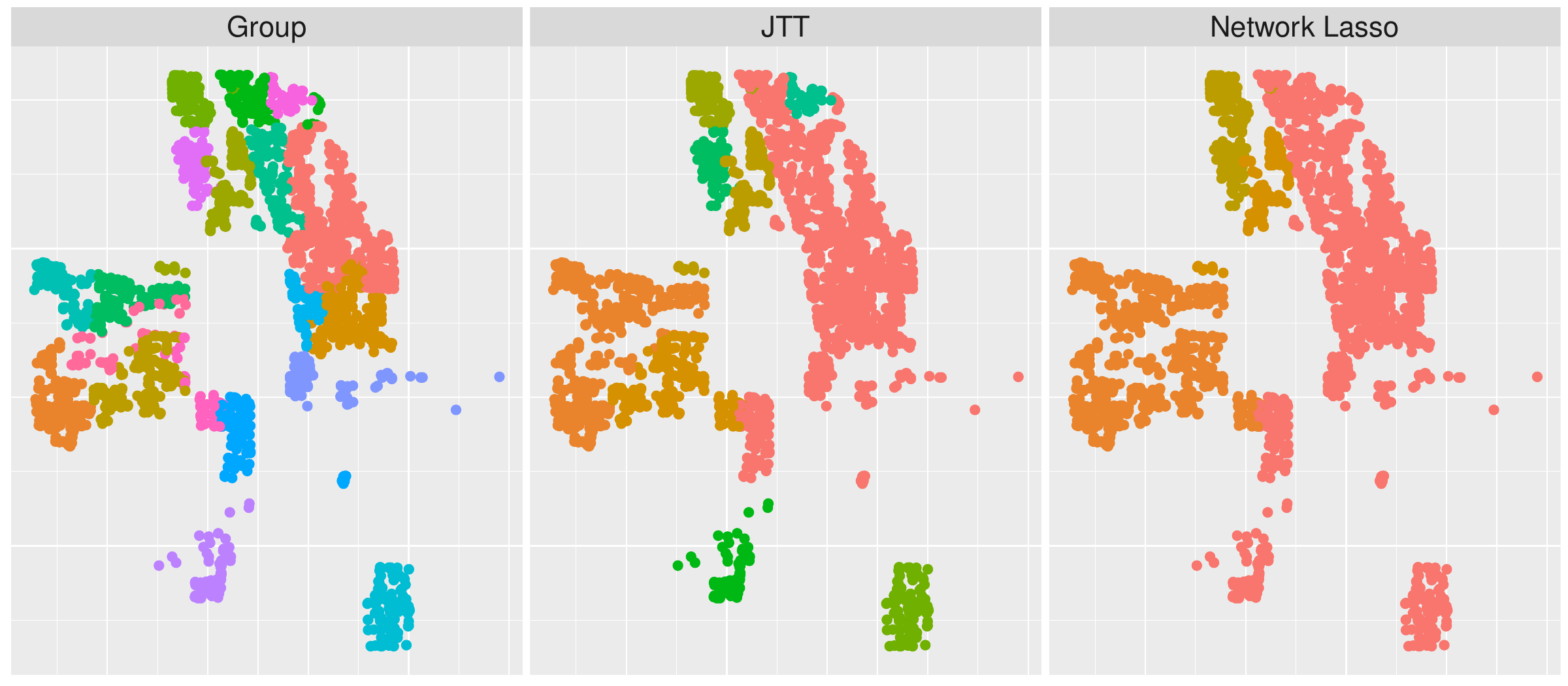}
  \caption{\ Initial 19 groups, and clustering results \label{fig clustering}}
\end{figure}
Figure~\ref{fig clustering} shows the initial 19 groups and the clustering results for the JTT method and network Lasso, which produced 9 and 4 clusters, respectively.
\begin{table}[h]
\caption{\ Results of leave-one-out cross-validation \label{tab LOOCV}}
\centering
\begin{tabular}{rrrrrrrrrrr}
\toprule
  &  &  & \multicolumn{8}{c}{Fitting} \\
\cmidrule(lr){4-11}
 \multicolumn{3}{c}{Prediction accuracy (USD)} & \multicolumn{3}{c}{$R^2$} & \multicolumn{2}{c}{Clusters} & \multicolumn{3}{c}{Runtime (sec.)} \\
\cmidrule(lr){1-3} \cmidrule(lr){4-6} \cmidrule(lr){7-8} \cmidrule(lr){9-11}
 JTT1 & JTT2 & NL & JTT1 & JTT2 & NL & JTT & NL & JTT1 & JTT2 & NL \\
\midrule
 30588.0 & 28669.1 & 36500.2 & 0.91 & 0.90 & 0.80 & 8.999 & 3.954 & 0.044 & 0.102 & 104.906 \\
\bottomrule
\end{tabular}
\end{table}
Table~\ref{tab LOOCV} summarizes the results of leave-one-out cross-validation, in which prediction accuracy is measured as $\sqrt{\sum_{i=1}^n (y_i^\star - \hat{y}_i^\star)^2/n}$, where $y_i^\star$ and $\hat{y}_i^\star$ are the $i$th test data and the predicted value, respectively, and other values are averages of the coefficients of determination, numbers of estimated clusters, and runtimes of fitting results for the training data.
From the table, we can see that the JTT method is much faster and has higher prediction accuracy than network Lasso.
Moreover, the penalized estimation method in JTT2 was able to improve prediction accuracy relative to JTT1.

\section{Conclusion}

For group-wise linear regression, we proposed the join-two-together (JTT) method for group clustering when the relationships of the groups are given by a graph.
By formulating the clustering problem as the edge selection problem in the graph, we were able to formulate a simple method based on the KOO method.
The performance of the JTT method depends on the model selection criterion used to evaluate the goodness of a model.
This paper presented a condition such that the JTT method based on a $GC_p$ criterion has consistency in the edge selection under a high-dimensional asymptotic framework, and proposed a specific criterion.
Furthermore, to implement the JTT method, R package \texttt{JTT} was developed.

Simulation demonstrated that the JTT method has consistency.
Furthermore, we found that a simple penalized estimation method can improve MSEs for the fitted values and the estimator of regression coefficients.
Moreover, the JTT method performed much better than network Lasso in terms clustering accuracy, MSE, and runtime, in both simulation and an application to real data.
Although network Lasso was implemented under the default settings of R package \texttt{GGFL}, its performance may be improved by changing the model selection criterion to select the tuning parameter and penalty weights.
However, since we cannot expect significant improvement in calculation speed in that way, the superiority of the JTT method can be considered invariant.
The present paper focused on group clustering, but selection of explanatory variables is also important.
By adding a penalty for variable selection, network Lasso can be extended  to perform group clustering and variable selection simultaneously.
However, needless to say, its optimization becomes more complex and it requires more computation time.
On the other hand, when using the JTT method to perform clustering, variable selection can be carried out using the KOO method.
In that case, consistent variable selection becomes possible with the number of calculations increasing by only $p$ (or  $mp$).
Furthermore, the KOO method does not require additional calculation of inverse matrices.
Hence, clustering and variable selection can be performed at high speed.

However, the JTT method requires a sufficiently large sample size for each group to guarantee consistency.
Similar to the Ames Housing Data utilized in Section~\ref{sec realdata}, it is plausible that some small sample groups appear in practical scenarios.
Therefore, we need to consider a method for coping with this problem.

\paragraph{Acknowledgements}

The authors thank Prof.\ Hirokazu Yanagihara of Osaka Metropolitan University for his many helpful comments and FORTE Science Communications (\url{https://www.forte-science.co.jp/}) for English language editing.
This work was partially supported by JSPS KAKENHI Grant Numbers 25K17296 and 25K21159.


\bibliographystyle{unsrtnat}
\bibliography{bibliography}  

\begin{thebibliography}{26}
\providecommand{\natexlab}[1]{#1}
\providecommand{\url}[1]{\texttt{#1}}
\expandafter\ifx\csname urlstyle\endcsname\relax
  \providecommand{\doi}[1]{doi: #1}\else
  \providecommand{\doi}{doi: \begingroup \urlstyle{rm}\Url}\fi

\bibitem[Hallac et~al.(2015)Hallac, Leskovec, and Boyd]{Hallac-2015}
D.~Hallac, J.~Leskovec, and S.~Boyd.
\newblock Network {L}asso: {C}lustering and optimization in large graphs.
\newblock In \emph{Proceedings of the 21th ACM SIGKDD International Conference
  on Knowledge Discovery and Data Mining}, KDD '15, pages 387--396, New York,
  NY, USA, 2015. Association for Computing Machinery.

\bibitem[Ohishi et~al.(2025)Ohishi, Okamura, Itoh, Wakaki, and
  Yanagihara]{Ohishi-2025}
M.~Ohishi, K.~Okamura, Y.~Itoh, H.~Wakaki, and H.~Yanagihara.
\newblock Coordinate descent algorithm for generalized group fused {L}asso.
\newblock \emph{Behaviormetrika}, 52:\penalty0 105--137, 2025.

\bibitem[Boyd et~al.(2011)Boyd, Parikh, Chu, Peleato, and Eckstein]{Boyd-2011}
S.~Boyd, N.~Parikh, E.~Chu, B.~Peleato, and J.~Eckstein.
\newblock Distributed optimization and statistical learning via the alternating
  direction method of multipliers.
\newblock \emph{Found. Trends Mach. Learn.}, 3:\penalty0 1--122, 2011.

\bibitem[Zhao et~al.(1986)Zhao, Krishnaiah, and Bai]{Zhao-1986}
L.~C. Zhao, P.~R. Krishnaiah, and Z.~D. Bai.
\newblock On detection of the number of signals in presence of white noise.
\newblock \emph{J. Multivariate Anal.}, 20:\penalty0 1--25, 1986.

\bibitem[Nishii et~al.(1988)Nishii, Bai, and Krishnaiah]{Nishii-1988}
R.~Nishii, Z.~D. Bai, and P.~R. Krishnaiah.
\newblock Strong consistency of the information criterion for model selection
  in multivariate analysis.
\newblock \emph{Hiroshima Math. J.}, 18:\penalty0 451--462, 1988.

\bibitem[Bai et~al.(2018)Bai, Fujikoshi, and Hu]{Bai-2018tr}
Z.~D. Bai, Y.~Fujikoshi, and J.~Hu.
\newblock Strong consistency of the {AIC}, {BIC}, {$C_p$} and {KOO} methods in
  high-dimensional multivariate linear regression.
\newblock Technical Report TR-No. 18-09, Hiroshima Statistical Research Group,
  Hiroshima, 2018.

\bibitem[Akaike(1973)]{Akaike1973}
H.~Akaike.
\newblock Information theory and an extension of the maximum likelihood
  principle.
\newblock In B.~N. Petrov and F.~Cs{\'{a}}ki, editors, \emph{2nd International
  Symposium on Information Theory}, pages 267--281, Budapest, 1973.
  Akad{\'{e}}miai Kiad{\'{o}}.

\bibitem[Mallows(1973)]{Mallows1973}
C.~L. Mallows.
\newblock Some comments on {$C_p$}.
\newblock \emph{Technometrics}, 15:\penalty0 661--675, 1973.

\bibitem[Oda and Yanagihara(2020)]{OdaYanagihara2020}
R.~Oda and H.~Yanagihara.
\newblock A fast and consistent variable selection method for high-dimensional
  multivariate linear regression with a large number of explanatory variables.
\newblock \emph{Electron. J. Stat.}, 14:\penalty0 1386--1412, 2020.

\bibitem[Oda and Yanagihara(2021)]{OdaYanagihara2021}
R.~Oda and H.~Yanagihara.
\newblock A consistent likelihood-based variable selection method in normal
  multivariate linear regression.
\newblock In I.~Czarnowski, R.~J. Howlett, and L.~C. Jain, editors,
  \emph{Intelligent Decision Technologies}, pages 391--401, Singapore, 2021.
  Springer Singapore.

\bibitem[Atkinson(1980)]{Atkinson1980}
A.~C. Atkinson.
\newblock A note on the generalized information criterion for choice of a
  model.
\newblock \emph{Biometrika}, 67:\penalty0 413--418, 1980.

\bibitem[Nishii(1984)]{Nishii1984}
R.~Nishii.
\newblock Asymptotic properties of criteria for selection of variables in
  multiple regression.
\newblock \emph{Ann. Statist.}, 12:\penalty0 758--765, 1984.

\bibitem[Wang and Leng(2008)]{WangLeng2008}
H.~Wang and C.~Leng.
\newblock A note on adaptive group {L}asso.
\newblock \emph{Comput. Statist. Data Anal.}, 52:\penalty0 5277--5286, 2008.

\bibitem[Yanagihara(2016)]{Yanagihara2016}
H.~Yanagihara.
\newblock A high-dimensionality-adjusted consistent {$C_p$}-type statistic for
  selecting variables in a normality-assumed linear regression with multiple
  responses.
\newblock \emph{Procedia Comput. Sci.}, 96:\penalty0 1096--1105, 2016.

\bibitem[Schwarz(1978)]{Schwarz1978}
G.~Schwarz.
\newblock Estimating the dimension of a model.
\newblock \emph{Ann. Statist.}, 6:\penalty0 461--464, 1978.

\bibitem[Fujikoshi and Satoh(1997)]{FujikoshiSatoh1997}
Y.~Fujikoshi and K.~Satoh.
\newblock Modified {AIC} and {$C_p$} in multivariate linear regression.
\newblock \emph{Biometrika}, 84:\penalty0 707--716, 1997.

\bibitem[Yanagihara(2018)]{Yanagihara2018}
H.~Yanagihara.
\newblock Explicit solution to the minimization problem of generalized
  cross-validation criterion for selecting ridge parameters in generalized
  ridge regression.
\newblock \emph{Hiroshima Math. J.}, 48:\penalty0 203--222, 2018.

\bibitem[{R Core Team}(2025)]{R450}
{R Core Team}.
\newblock \emph{R: A Language and Environment for Statistical Computing}.
\newblock R Foundation for Statistical Computing, Vienna, Austria, 2025.
\newblock URL \url{https://www.R-project.org/}.

\bibitem[Ohishi(2025{\natexlab{a}})]{GGFL102}
M.~Ohishi.
\newblock \emph{GGFL: Coordinate optimization for GGFL}, 2025{\natexlab{a}}.
\newblock URL \url{https://github.com/ohishim/GGFL}.
\newblock {R} package version 1.0.2.

\bibitem[Ohishi(2025{\natexlab{b}})]{JTT010}
M.~Ohishi.
\newblock \emph{JTT: Consistent clustering for group-wise linear regression},
  2025{\natexlab{b}}.
\newblock URL \url{https://github.com/ohishim/JTT}.
\newblock {R} package version 0.1.0.

\bibitem[Ohishi et~al.(2021)Ohishi, Fukui, Okamura, Itoh, and
  Yanagihara]{Ohishi-2021}
M.~Ohishi, K.~Fukui, K.~Okamura, Y.~Itoh, and H.~Yanagihara.
\newblock Coordinate optimization for generalized fused {L}asso.
\newblock \emph{Comm. Statist. Theory Methods}, 50:\penalty0 5955--5973, 2021.

\bibitem[Ohishi et~al.(2020)Ohishi, Yanagihara, and Fujikoshi]{Ohishi-2020}
M.~Ohishi, H.~Yanagihara, and Y.~Fujikoshi.
\newblock A fast algorithm for optimizing ridge parameters in a generalized
  ridge regression by minimizing a model selection criterion.
\newblock \emph{J. Statist. Plann. Inference}, 204:\penalty0 187--205, 2020.
\newblock \doi{10.1016/j.jspi.2019.04.010}.

\bibitem[Kuhn(2025)]{modeldata151}
M.~Kuhn.
\newblock \emph{modeldata: Data sets useful for modeling examples}, 2025.
\newblock URL \url{https://CRAN.R-project.org/package=modeldata}.
\newblock {R} package version 1.5.1.

\bibitem[De~Cock(2011)]{DeCock2011}
D.~De~Cock.
\newblock {A}mes, {I}owa: {A}lternative to the {B}oston housing data as an end
  of semester regression project.
\newblock \emph{J. Stat. Educ.}, 19, 2011.

\bibitem[Yanagihara(2024)]{Yanagihara2024}
H.~Yanagihara.
\newblock High-dimensionality-adjusted consistent {AIC} in normal multivariate
  linear regression.
\newblock \emph{Procedia Comput. Sci.}, 246:\penalty0 2022--2031, 2024.

\bibitem[Imori et~al.(2014)Imori, Katayama, and Wakaki]{Imori-2014tr}
S.~Imori, S.~Katayama, and H.~Wakaki.
\newblock Screening and selection methods in high-dimensional linear regression
  model.
\newblock Technical Report TR-No. 14-01, Hiroshima Statistical Research Group,
  Hiroshima, 2014.

\end{thebibliography}






\appendix
\section*{Appendix}
\setcounter{section}{1}
\setcounter{equation}{0}
\renewcommand{\theequation}{{\rm A}.\arabic{equation}}
\numberwithin{Lem}{section}

\subsection{The proof of equation \eqref{delmin}}
\label{ap delmin}

Let $\vA_{k \ell} = \vM_k \vM_{k \ell}^{-1} \vM_{\ell}$.
Since $\vM_{k \ell} = \vM_k + \vM_\ell$, using the identity $\vA - \vA (\vA + \vB)^{-1} \vA = \vA (\vA + \vB)^{-1} \vB$ for nonsingular matrices $\vA$ and $\vB$, the coefficient matrix of the quadratic form $\delta_{k \ell}\ ((k, \ell) \notin E_\ast)$ for $({\vbeta_k^\ast}', {\vbeta_\ell^\ast}')'$ can be expressed as
\begin{align*}
  &\begin{pmatrix}
    \vX_k' & \vO_{p, n_\ell} \\ \vO_{p, n_k} & \vX_\ell'
  \end{pmatrix} (\vI_{n_{k \ell}} - \vP_{k \ell}) \begin{pmatrix}
    \vX_k & \vO_{n_k, p} \\ \vO_{n_\ell, p} & \vX_\ell
  \end{pmatrix} \\
  &\qquad\quad = \begin{pmatrix}
      \vM_k - \vM_k \vM_{k \ell}^{-1} \vM_k & - \vM_k \vM_{k \ell}^{-1} \vM_\ell \\
      - \vM_\ell \vM_{k \ell}^{-1} \vM_k & \vM_\ell - \vM_\ell \vM_{k \ell}^{-1} \vM_\ell
    \end{pmatrix}
  = \begin{pmatrix}
      \vA_{k \ell} & - \vA_{k \ell} \\ - \vA_{k \ell} & \vA_{k \ell}
     \end{pmatrix},
\end{align*}
where $\vO_{n, p}$ is an $n \times p$ matrix of zeros.
Since $\vA_{k \ell}$ is a symmetric matrix, for $\vgamma_{k \ell} = \vbeta_k^\ast - \vbeta_\ell^\ast$, we have
\begin{align*}
  \sigma_\ast^2 \delta_{k \ell}
    &= \left( {\vbeta_k^\ast}', {\vbeta_\ell^\ast}' \right) \begin{pmatrix}
           \vA_{k \ell} & - \vA_{k \ell} \\ - \vA_{k \ell} & \vA_{k \ell}
         \end{pmatrix} \begin{pmatrix}
           \vbeta_k^\ast \\ \vbeta_\ell^\ast
         \end{pmatrix}
    = {\vbeta_k^\ast}' \vA_{k \ell} \vbeta_k^\ast - 2 {\vbeta_k^\ast}' \vA_{k \ell} \vbeta_\ell^\ast + {\vbeta_\ell^\ast}' \vA_{k \ell} \vbeta_\ell^\ast
    = \vgamma_{k \ell}' \vA_{k \ell} \vgamma_{k \ell} \\
    &\ge \lambda_{\min} (\vA_{k \ell}) \| \vgamma_{k \ell} \|^2.
\end{align*}
Hence, Assumptions A2 and A3 imply
\begin{align*}
  n_0^{-c_3} \delta_{\min}
    &\ge n_0^{-c_3} \min_{(k, \ell) \notin E_\ast} \lambda_{\min} (\vA_{k \ell}) \| \vgamma_{k \ell} \|^2 / \sigma_\ast^2 \\
    &\ge \left\{ n_0^{-1} \min_{(k, \ell) \notin E_\ast} \lambda_{\min} (\vA_{k \ell}) \right\} \left\{
            n_0^{1-c_3} \min_{(k, \ell) \notin E_\ast} \| \vgamma_{k \ell} \|^2 / \sigma_\ast^2
          \right\} 
    \ge c_1 c_2,
\end{align*}
and \eqref{delmin} is proved.

\subsection{The proof of Theorem~\ref{th main}}
\label{ap main}

We prove the theorem with the following lemma given in \cite{OdaYanagihara2020}.
\begin{Lem} {\it
\label{lem A1}
  Let $u_1$, $u_2$, and $v$ be random variables distributed according to $\chi^2 (p)$, $\chi^2 (p, \delta)$, and $\chi^2 (N)$, respectively, where $u_1$ and $u_2$ are independent of $v$ and $N = n - mp$.
  Then, for $N - 4 r > 0\ (r \in \bN)$, the following expressions are true for $N \to \infty$.
  \begin{align*}
    \E \left[ \left(
      \dfrac{u_1}{v} - \dfrac{p}{N-2}
    \right)^{2r} \right] &= O (p^r n^{-2 r}), \\
    \E \left[ \left(
      \dfrac{u_2}{v} - \dfrac{p + \delta}{N - 2}
    \right)^{2r} \right] &= O \left( \max \left\{
      (p + \delta)^r n^{-2 r}, (p + \delta)^{2r} n^{-3 r}
    \right\} \right).
  \end{align*}
}\end{Lem}

From \eqref{Ehat}, the probability of $\hat{E} (\alpha) = E_\ast$ can be evaluated as follows:
\begin{align*}
  \Pr \left( \hat{E} (\alpha) = E_\ast \right)
    &= \Pr \left(
           \left\{
             \bigcap_{(k, \ell) \in E_\ast} \left\{ 
               GC_p^{(k \ell)} (\alpha) \le GC_p^{(0)} (\alpha) 
             \right\}
           \right\} \bigcap \left\{
             \bigcap_{(k, \ell) \notin E_\ast} \left\{ 
               GC_p^{(k \ell)} (\alpha) > GC_p^{(0)} (\alpha) 
             \right\}
           \right\}
         \right) \\
    &\ge 1 - \sum_{(k, \ell) \in E_\ast} \Pr \left(
              GC_p^{(k \ell)} (\alpha) > GC_p^{(0)} (\alpha)
            \right) - \sum_{(k, \ell) \notin E_\ast} \Pr \left(
              GC_p^{(k \ell)} (\alpha) < GC_p^{(0)} (\alpha)
            \right).
\end{align*}
Hence, under the asymptotic framework in \eqref{AFW}, it is sufficient to show that
\begin{align}
\label{ap eq1}
  \sum_{(k, \ell) \in E_\ast} \Pr \left(GC_p^{(k \ell)} (\alpha) > GC_p^{(0)} (\alpha) \right)
    = o (1), \quad
  \sum_{(k, \ell) \notin E_\ast} \Pr \left(GC_p^{(k \ell)} (\alpha) < GC_p^{(0)} (\alpha) \right)
    = o (1).
\end{align}

The difference of $GC_p$ criteria in \eqref{GCp} is given by
\begin{align*}
  GC_p^{(k \ell)} (\alpha) - GC_p^{(0)} (\alpha)
    &= \dfrac{1}{s^2} \left\{
           \vy_{k \ell}' (\vI_{n_{k\ell}} - \vP_{k \ell}) \vy_{k \ell}
             - \sum_{j \notin V_{k \ell}} \vy_j' (\vI_{n_j} - \vP_j) \vy_j
         \right\} - \alpha p \\
    &= N \dfrac{\vy_{k \ell}' \{ \diag (\vP_k, \vP_\ell) - \vP_{k \ell} \} \vy_{k \ell}}{\sum_{j=1}^m \vy_j' (\vI_{n_j} - \vP_j) \vy_j} - \alpha p,
\end{align*}
where $\diag (\vA, \vB)$ indicates the block diagonal matrix with $\vA$ and $\vB$ as the diagonal blocks.
Let $u_{k \ell} = \vy_{k \ell}' \{ \diag (\vP_k, \vP_\ell) - \vP_{k \ell} \} \vy_{k \ell} / \sigma_\ast^2$ and $v = \sum_{j=1}^m \vy_j' (\vI_{n_j} - \vP_j) \vy_j / \sigma_\ast^2$.
Then $u_{k \ell}$ and $v$ can be rewritten as quadratic forms for $\vy = (\vy_1', \ldots, \vy_m')'$.
Their coefficient matrices are idempotent and their product is $\vO_{n, n}$.
Hence, Cochran's theorem gives that $u_{k \ell}$ and $v$ are independent and that $u_{k \ell} \sim \chi^2 (p, \delta_{k \ell})$ and $v \sim \chi^2 (N)$.
Recall that $\delta_{k \ell} = 0$ holds when $(k, \ell) \in E_\ast$.
Hence, we have
\begin{align}
\label{difGCp}
  GC_p^{(k \ell)} (\alpha) - GC_p^{(0)} (\alpha) = \begin{dcases}
      \dfrac{u_0 N}{v} - \alpha p
        & ((k, \ell) \in E_\ast) \\
      \dfrac{u_{k \ell} N}{v} - \alpha p
        & ((k, \ell) \notin E_\ast)
    \end{dcases},
\end{align}
where $u_0$ is independent of $v$ and $u_0 \sim \chi^2 (p)$.

We first show \eqref{ap eq1} when $(k, \ell) \in E_\ast$.
Regarding the first equation in \eqref{ap eq1}, we have
\begin{align*}
  \sum_{(k, \ell) \in E_\ast} \Pr \left(GC_p^{(k \ell)} (\alpha) > GC_p^{(0)} (\alpha) \right)
    &= q_\ast \Pr \left( \dfrac{u_0}{v} > \dfrac{\alpha p}{N} \right)
    = q_\ast \Pr \left( \dfrac{u_0}{v} - \dfrac{p}{N - 2} > \rho \right) \\
    &\le q_\ast \Pr \left( \left| \dfrac{u_0}{v} - \dfrac{p}{N - 2} \right| \ge \rho \right),
\end{align*}
where $q_\ast = \# (E_\ast)$, $\rho = \beta p / N$, and $\beta = \alpha - N / (N - 2)$.
For $r_1 \in \bN$, it holds from Markov's inequality and Lemma~\ref{lem A1} that
\begin{align*}
  \Pr \left( \left| \dfrac{u_0}{v} - \dfrac{p}{N - 2} \right| \ge \rho \right)
    &\le \rho^{-2 r_1} \E \left[ \left(
             \dfrac{u_0}{v} - \dfrac{p}{N - 2}
           \right)^{2 r_1} \right]
    = O (\beta^{-2r_1} p^{-r_1}).
\end{align*}
Since $q_\ast = O (m^2)$ and $\beta p^{1/2} / m^{1/r_1} \to \infty$, we have
\begin{align}
\label{ap over1}
  \sum_{(k, \ell) \in E_\ast} \Pr \left(GC_p^{(k \ell)} (\alpha) > GC_p^{(0)} (\alpha) \right)
    = O (m^2 \beta^{-2r_1} p^{-r_1})
    = o (1).
\end{align}

Regarding the second equation in \eqref{ap eq1}, when $(k, \ell) \notin E_\ast$, we have
\begin{align*}
  \sum_{(k, \ell) \notin E_\ast} \Pr \left(GC_p^{(k \ell)} (\alpha) < GC_p^{(0)} (\alpha) \right)
    &= \sum_{(k, \ell) \notin E_\ast} \Pr \left( \dfrac{u_{k \ell}}{v} < \dfrac{\alpha p}{N} \right)
    = \sum_{(k, \ell) \notin E_\ast} \Pr \left( 
         \dfrac{u_{k \ell}}{v} - \dfrac{p + \delta_{k \ell}}{N - 2} < \rho - \dfrac{\delta_{k \ell}}{N - 2}
       \right).
\end{align*}
Since \eqref{delmin} and $\beta p / n_0^{c_3} \to 0$ imply $\rho - \delta_{k \ell} / (N - 2) < 0$ for sufficiently large $N$, we have
\begin{align*}
  \Pr \left( 
    \dfrac{u_{k \ell}}{v} - \dfrac{p + \delta_{k \ell}}{N - 2} < \rho - \dfrac{\delta_{k \ell}}{N - 2}
  \right) \le \Pr \left( 
    \left| \dfrac{u_{k \ell}}{v} - \dfrac{p + \delta_{k \ell}}{N - 2} \right| \ge \dfrac{\delta_{k \ell}}{N - 2} - \rho
  \right).
\end{align*}
Furthermore, for $r_2 \in \bN$, it holds from Markov's inequality and Lemma~\ref{lem A1} that
\begin{align*}
  \sum_{(k, \ell) \notin E_\ast} \Pr \left( 
    \left| \dfrac{u_{k \ell}}{v} - \dfrac{p + \delta_{k \ell}}{N - 2} \right| \ge \dfrac{\delta_{k \ell}}{N - 2} - \rho
  \right)
    &\le \sum_{(k, \ell) \notin E_\ast} \left( \dfrac{\delta_{k \ell}}{N - 2} - \rho \right)^{-2 r_2} \E \left[ \left(
             \dfrac{u_{k \ell}}{v} - \dfrac{p + \delta_{k \ell}}{N - 2}
           \right)^{2 r_2} \right] \\
    &\le (q - q_\ast) \max_{(k, \ell) \notin E_\ast} \left( \dfrac{\delta_{k \ell}}{N - 2} - \rho \right)^{-2 r_2} \E \left[ \left(
             \dfrac{u_{k \ell}}{v} - \dfrac{p + \delta_{k \ell}}{N - 2}
           \right)^{2 r_2} \right], \\
  \E \left[ \left(
    \dfrac{u_{k \ell}}{v} - \dfrac{p + \delta_{k \ell}}{N - 2}
  \right)^{2 r_2} \right] 
    &= O \left( \max \left\{
           \dfrac{(p + \delta_{k \ell})^{r_2}}{n^{2 r_2}}, \dfrac{(p + \delta_{k \ell})^{2 r_2}}{n^{3 r_2}}
         \right\} \right).
\end{align*}
Notice that the following inequalities hold for sufficiently large $N$:
\begin{align*}
  \left( \dfrac{N - 2}{\delta_{k \ell}} \right)^{2 r_2} \left( \dfrac{p + \delta_{k \ell}}{n^2} \right)^{r_2}
    \le \left( \dfrac{1 + p / \delta_{\min}}{\delta_{\min}} \right)^{r_2}, \quad
  \left( \dfrac{N - 2}{\delta_{k \ell}} \right)^{2 r_2} \left( \dfrac{p + \delta_{k \ell}}{n^{3/2}} \right)^{2 r_2}
    \le \left( \dfrac{1 + p / \delta_{\min}}{n^{1/2}} \right)^{2 r_2}.
\end{align*}
Hence, we have
\begin{align}
  &(q - q_\ast) \max_{(k, \ell) \notin E_\ast} \left( \dfrac{\delta_{k \ell}}{N - 2} - \rho \right)^{-2 r_2} \left\{
    \dfrac{(p + \delta_{k \ell})^{r_2}}{n^{2 r_2}} + \dfrac{(p + \delta_{k \ell})^{2 r_2}}{n^{3 r_2}}
  \right\} \nonumber \\
  &\qquad\quad \le (q - q_\ast) \left( 1 - \dfrac{\rho (N - 2)}{\delta_{\min}} \right)^{-2 r_2} \left\{
                              \left( \dfrac{1 + p / \delta_{\min}}{\delta_{\min}} \right)^{r_2}
                                + \left( \dfrac{1 + p / \delta_{\min}}{n^{1/2}} \right)^{2 r_2}
                            \right\} \nonumber \\
  &\qquad\quad = O \left( m^2 \delta_{\min}^{-r_2} \right) + O \left( m^2 p^{r_2} \delta_{\min}^{-2r'} \right)
                           + O \left( m^2 n^{-r_2} \right) + O \left( m^2 p^{2 r_2} n^{-r_2} \delta_{\min}^{-2r'} \right) \nonumber \\
  &\qquad \quad = O \left( 
                             n_0^{2 - c_3 r_2} + n_0^{2 + r_2 - 2 c_3 r_2}
                               + m^{2 - r_2} n_0^{-r_2} + m^{2 - r_2} n_0^{r_2 - 2 c_3 r_2}
                           \right). \label{ap under1}
\end{align}
If $r_2 > 2$ and $c_3 > (r_2 + 2) / 2 r_2$, then \eqref{ap under1} converges to 0.
Since $r_2$ is arbitrary and $(r_2 + 2) / 2 r_2 > 1/2$ holds for $r_2 > 2$, it follows that $c_3 > (r_2 + 2) / 2 r_2$ is equivalent to $c_3 \ge 1/2$.
Hence, we have
\begin{align*}
  \sum_{(k, \ell) \notin E_\ast} \Pr \left(GC_p^{(k \ell)} (\alpha) < GC_p^{(0)} (\alpha) \right)
    = o (1).
\end{align*}

As discussed above, \eqref{ap eq1} holds and the probability of $\hat{E} (\alpha) = E_\ast$ converges to 1.
Moreover, we have the convergence order from \eqref{ap over1} and \eqref{ap under1}, and consequently Theorem~\ref{th main} is proved.

\subsection{The detail of $B$ in \eqref{alpha}}
\label{ap B}

Based on \cite{Yanagihara2024}, we explain how $B$, which is included in $\hat{\beta}$ in \eqref{alpha}, standardizes the differences of $HCGC_p$ criteria.
Based on \eqref{difGCp}, we define $D_{k \ell} (\alpha)$ as
\begin{align*}
  D_{k \ell} (\alpha)
    = \dfrac{u_{k \ell}}{v}
    = \dfrac{1}{N} \left\{ HCGC_p^{(k \ell)} (\alpha) - HCGC_p^{(0)} (\alpha) + \alpha p \right\}.
\end{align*}
Notice that $\E [D_{k \ell} (\hat{\alpha})] = \mu = p / (N - 2)$ and $\Var [D_{k \ell} (\hat{\alpha})] = \tau^2 = 2 p B^2 / N^2$ from $u_{k \ell} \sim \chi^2 (p)$ and $v \sim \chi^2 (N)$ when $(k, \ell) \in E_\ast$.
Hence, we have $N^{-1} \{ HCGC_p^{(k \ell)} (\hat{\alpha}) - HCGC_p^{(0)} (\hat{\alpha}) \} = D_{k \ell} (\hat{\alpha}) - \mu - \tau m^{1/4} \log n_0 / \sqrt{2}$.
This expression implies
\begin{align*}
  \Pr \left( HCGC_p^{(k \ell)} (\hat{\alpha}) - HCGC_p^{(0)} (\hat{\alpha}) > 0 \right)
    = \Pr \left( \dfrac{D_{k \ell} (\hat{\alpha}) - \mu}{\tau} > \dfrac{m^{1/4} \log n_0}{\sqrt{2}} \right),
\end{align*}
and we can see that $B$ standardizes the difference of $HCGC_p$ criteria with $\alpha = \hat{\alpha}$.

\subsection{The proof of Theorem~\ref{th sub1}}
\label{ap sub1}

We use the following lemma to prove the theorem (e.g., \citealp{Imori-2014tr}).
\begin{Lem} {\it
\label{lem A2}
  Let $z$, $u_1$, and $u_2$ be random variables distributed according to $N (0, 1)$, $\chi^2 (p)$, and $\chi^2 (p, \delta)$, respectively.
  Then, we have
  \begin{align*}
    \Pr (|z| > r) 
      &\le \exp (- r^2 / 2), \\
    \Pr (u_1 > (1 + r) p) 
      &\le \exp (- h r p) \quad (r \in [1, \infty),\ h = (1 - \log 2) / 2), \\
    \Pr (u_1 < (1 - r) p)
      &\le \exp (- r p / 4) \quad (r \in [0, 1)), \\
    \Pr (u_2 < r)
      &\le \Pr (|z| > t) + \Pr (u_1 < r - \delta + 2 \sqrt{\delta} t) \quad (t \ge 0).
  \end{align*}
}\end{Lem}

Similar to the proof of Theorem~\ref{th main}, it is sufficient to show \eqref{ap eq1}.
We first evaluate the probability for $(k, \ell) \in E_\ast$ for $r_1 \in (0, 1)$ as
\begin{align}
\label{ap eq3}
  \sum_{(k, \ell) \in E_\ast} \Pr \left(GC_p^{(k \ell)} (\alpha) > GC_p^{(0)} (\alpha) \right)
    &= q_\ast \Pr \left( \dfrac{u_0}{v} > \dfrac{(1 - r_1) \alpha p}{(1 - r_1) N} \right) \nonumber \\
    &\le q_\ast \Pr (u_0 > (1 - r_1) \alpha p) + q_\ast \Pr (v < (1 - r_1) N).
\end{align}
Regarding the first term in \eqref{ap eq3}, since $(1 - r_1) \alpha = 1 + \{ (1 - r_1) \alpha - 1 \}$, it holds from Lemma~\ref{lem A2} that
\begin{align*}
  q_\ast \Pr (u_0 > (1 - r_1) \alpha p)
    &\le q_\ast \exp [- h \{ (1 - r_1) \alpha - 1 \} p]
    = O (m^2 \exp [- h \{ (1 - r_1) \alpha - 1 \} p]),
\end{align*}
for $\alpha \ge 2 / (1 - r_1)$.
Hence, we have
\begin{align*}
  m^2 \exp [- h \{ (1 - r_1) \alpha - 1 \} p]
    = \exp \left(
        - \alpha p \left[ 
          - \dfrac{2 \log m}{\alpha p} + h \left\{ (1 - r_1) - \dfrac{1}{\alpha} \right\}
        \right]
       \right)
    = o (1),
\end{align*}
from $(1 - r_1) - 1/\alpha > 0$ and $\alpha p / \log m \to \infty$.
On the other hand, regarding the second term in \eqref{ap eq3}, it holds from Lemma~\ref{lem A2} that
\begin{align*}
   q_\ast \Pr (v < (1 - r_1) N)
     &\le q_\ast \exp (- r_1 N / 4)
     = O \left( m^2 \exp (-r_1 N / 4) \right)
     = o (1).
\end{align*}
Hence, we have
\begin{align*}
  \sum_{(k, \ell) \in E_\ast} \Pr \left(GC_p^{(k \ell)} (\alpha) > GC_p^{(0)} (\alpha) \right)
    = o (1).
\end{align*}

Looking at $(k, \ell) \notin E_\ast$, for $r_2 \in [1, \infty)$, we have
\begin{align}
\label{ap eq4}
  \sum_{(k, \ell) \notin E_\ast} \Pr \left(GC_p^{(k \ell)} (\alpha) < GC_p^{(0)} (\alpha) \right)
    \le \sum_{(k, \ell) \notin E_\ast} \Pr (u_{k \ell} < (1 + r_2) \alpha p) + (q - q_\ast) \Pr (v > (1 + r_2) N).
\end{align}
Regarding the first term of the RHS in the above expression, for $z \sim N (0, 1)$ and $t\ (\ge 0)$, it holds from Lemma~\ref{lem A2} that
\begin{align*}
  \Pr (u_{k \ell} < (1 + r_2) \alpha p)
    \le \Pr (|z| > t) + \Pr \left( u_0 \le (1 + r_2) \alpha p - \delta_{k \ell} + 2 \sqrt{\delta_{k \ell}} t \right).
\end{align*}
Since $\alpha p / n_0^{c_3} \to 0$, \eqref{delmin} and Lemma~\ref{lem A2}, for $t = \{ c_1 c_2 n_0^{c_3} - (1 + r_2) \alpha p \} / 2 \sqrt{c_1 c_2 n_0^{c_3}}$ and sufficiently large $n_0$, we have
\begin{align*}
  \sum_{(k, \ell) \notin E_\ast} \Pr (u_{k \ell} < (1 + r_2) \alpha p)
    &\le \sum_{(k, \ell) \notin E_\ast} \Pr (|z| > t)
    \le \sum_{(k, \ell) \notin E_\ast} \exp \left( - t^2 \right) \\
    &= (q - q_\ast) \exp \left[
           - \dfrac{c_1 c_2 n_0^{c_3}}{8} \left\{ 1 - \dfrac{(1 + r_2) \alpha p}{c_1 c_2 n_0^{c_3}} \right\}^2
         \right] \\
    &= O \left( m^2 \exp (- c_1 c_2 n_0^{c_3} / 8) \right)
    = o (1).
\end{align*}
On the other hand, regarding the second term of the RHS in \eqref{ap eq4}, it holds from Lemma~\ref{lem A2} that
\begin{align*}
  (q - q_\ast) \Pr (v > (1 + r_2) N)
    \le (q - q_\ast) \exp (- h r_2 N)
    = O (m^2 \exp (- h r_2 N))
    = o (1).
\end{align*}
Hence, we have
\begin{align*}
  \sum_{(k, \ell) \notin E_\ast} \Pr \left(GC_p^{(k \ell)} (\alpha) < GC_p^{(0)} (\alpha) \right)
    = o (1).
\end{align*}
Consequently, \eqref{ap eq1} holds and Theorem~\ref{th sub1} is proved.

\subsection{Proof of Theorem~\ref{th sub2}}
\label{ap sub2}

We define $(k_1, \ell_1) \in E_\ast$ and $(k_2, \ell_2) \notin E_\ast$ such that, respectively, 
\begin{align*}
  &(k_1, \ell_1) \in E_\ast\ s.t.\ 
    \Pr \left( GC_p^{(k_1 \ell_1)} (\alpha) < GC_p^{(0)} (\alpha) \right) \not\to 1, \\
  &(k_2, \ell_2) \notin E_\ast\ s.t.\ 
    \Pr \left( GC_p^{(k_2 \ell_2)} (\alpha) > GC_p^{(0)} (\alpha) \right) \not\to 1.
 \end{align*}
Then, we have
\begin{align*}
  \Pr \left( \hat{E} (\alpha) = E_\ast \right)
    &= \Pr \left(
           \left\{
             \bigcap_{(k, \ell) \in E_\ast} \left\{ 
               GC_p^{(k \ell)} (\alpha) \le GC_p^{(0)} (\alpha) 
             \right\}
           \right\} \bigcap \left\{
             \bigcap_{(k, \ell) \notin E_\ast} \left\{ 
               GC_p^{(k \ell)} (\alpha) > GC_p^{(0)} (\alpha) 
             \right\}
           \right\}
         \right) \\
    &\le \left\{ \begin{array}{c}
             \Pr \left( GC_p^{(k_1 \ell_1)} (\alpha) < GC_p^{(0)} (\alpha) \right) \\
             \Pr \left( GC_p^{(k_2 \ell_2)} (\alpha) > GC_p^{(0)} (\alpha) \right)
           \end{array} \right\}
    \not\to 1.
\end{align*}
Hence, it is sufficient to show that either $(k_1, \ell_1)$ or $(k_2, \ell_2)$ exists.

From \eqref{difGCp}, for all $(k, \ell) \in E_\ast$, we have
\begin{align*}
  \Pr \left( GC_p^{(k \ell)} (\alpha) < GC_p^{(0)} (\alpha) \right)
    = \Pr \left( \dfrac{u_0}{v} < \dfrac{\alpha p}{N} \right).
\end{align*}
Notice that $v/N$ converges to 1 in probability, and $(u_0/p) / (v/N)$ also converges to 1 in probability as $p \to \infty$.
Hence, we have
\begin{align*}
\begin{dcases}
  \text{$p$: fixed and $\alpha \not\rightarrow \infty$}
    \Longrightarrow \Pr \left( \dfrac{u_0}{v} < \dfrac{\alpha p}{N} \right)
      = \Pr \left( \dfrac{u_0}{v/N} < \alpha p \right)
      \not\to 1 \\
  \text{$p \to \infty$ and $\lim \inf \alpha < 1$}
    \Longrightarrow \Pr \left( \dfrac{u_0}{v} < \dfrac{\alpha p}{N} \right)
      = \Pr \left( \dfrac{u_0/p}{v/N} < \alpha \right)
      \not\to 1
\end{dcases}.
\end{align*}

Let $(k_2, \ell_2) = \arg \min_{(k, \ell) \notin E_\ast} \delta_{k \ell}$.
Then we have
\begin{align*}
  \Pr \left( GC_p^{(k_2 \ell_2)} (\alpha) > GC_p^{(0)} (\alpha) \right)
    = \Pr \left( \dfrac{u_{k_2 \ell_2}}{v} > \dfrac{\alpha p}{N} \right).
\end{align*}
When $\delta_{\min} / p \to \infty$, it holds from Markov's inequality and $\alpha p / \delta_{\min} \to \infty$ that
\begin{align*}
  \Pr \left( \dfrac{u_{k_2 \ell_2}}{v} > \dfrac{\alpha p}{N} \right)
    \le \dfrac{N}{\alpha p} \cdot \dfrac{p + \delta_{\min}}{N - 2}
    = O (\delta_{\min} / \alpha p)
    = o (1).
\end{align*}
On the other hand, when $\delta_{\min} / p \to c_4\ (\in [0, \infty))$, since $\E [u_{k_2 \ell_2} / p] = 1 + \delta_{\min} / p \to 1 + c_4$ and $\Var [u_{k_2 \ell_2} / p] = 2 / p + 4 \delta_{\min} / p^2 \to 0$, $u_{k_2 \ell_2} / p$ converges to $1 + c_4$ in probability.
Hence, since $v / N$ converges to 1 in probability and $\lim \sup \alpha > 1 + c_4$, we have
\begin{align*}
  \Pr \left( \dfrac{u_{k_2 \ell_2}}{v} > \dfrac{\alpha p}{N} \right)
    = \Pr \left( \dfrac{u_{k_2 \ell_2} / p}{v / N} > \alpha \right)
    \not\to 1.
\end{align*}
Consequently, Theorem~\ref{th sub2} is proved.

\end{document}